\documentclass[]{article}
\usepackage[textwidth=17cm,textheight=25cm]{geometry}
\usepackage[utf8]{inputenc}
\usepackage{color}
\usepackage{scalerel}

\usepackage[
backend=biber,
style=phys,
sorting=none,
doi=false,
isbn=false,
arxiv=false
]{biblatex}

\renewbibmacro{in:}{}

\AtEveryBibitem{
	\clearfield{note}
	\clearfield{doi}
	\clearfield{urldate}
	\clearfield{urlyear}
	\clearfield{urlmonth}
	\clearfield{review}
	\clearfield{series}
	\clearlist{language}
}

\newcommand{\EMCT}{E_\text{\tiny MCT}}
\newcommand{\Tonset}{T_\text{onset}}
\newcommand{\TMCT}{T_\text{\tiny MCT}}
\newcommand{\TK}{T_\text{\tiny K}}
\newcommand{\TSF}{T_\text{\tiny SF}}
\newcommand{\IS}{\text{\tiny IS}}

\usepackage{amsmath}
\usepackage{amssymb} 
\usepackage{mathtools}
\usepackage[hidelinks]{hyperref}

\definecolor{cadmiumgreen}{rgb}{0.0, 0.42, 0.24}

\usepackage[symbol]{footmisc}

\date{October 21, 2020}

\title{Gradient descent dynamics in the mixed $p$-spin spherical model:\\
	finite size simulations and comparison with mean-field integration}

\usepackage{authblk}

\author[,1,2,3]{Giampaolo Folena\thanks{\texttt{giampaolo.folena@phys.ens.fr}}}
\author[1,4]{Silvio Franz}
\author[1,5]{Federico Ricci-Tersenghi}

\affil[1]{\footnotesize \textit{Dipartimento di Fisica, Universit\`{a} ``La Sapienza'', P.le A. Moro 5, 00185, Rome, Italy}}

\affil[2]{\footnotesize \textit{Laboratoire de Physique de l’Ecole Normale Sup\'erieure, ENS, Universit\'e PSL,}
	
	\textit{CNRS, Sorbonne Universit\'e, Universit\'e de Paris, F-75005 Paris, France}}
\affil[3]{ \footnotesize \textit{James Franck Institute and Department of Physics, University of Chicago, Chicago, IL 60637, U.S.A.}}

\affil[4]{\footnotesize \textit{LPTMS, UMR 8626, CNRS, Univ. Paris-Sud, Universit\'e Paris-Saclay, 91405 Orsay, France}}

\affil[5]{\footnotesize \textit{INFN, Sezione di Roma1, and CNR--Nanotec, Rome unit, P.le A. Moro 5, 00185, Rome, Italy}}

\begin{document}
	
\maketitle
	
\begin{abstract}
		
We perform numerical simulations of a long-range spherical spin glass with two and three body interaction terms. We study the gradient descent dynamics and the inherent structures found after a quench from initial conditions well thermalized at temperature $T_{in}$. In very large systems, the dynamics perfectly agrees with the integration of the mean-field dynamical equations. In particular, we confirm the existence of an onset initial temperature, within the liquid phase, below which the energy of the inherent structures undoubtedly depends on $T_{in}$. This behavior is in contrast with that of pure models, where there is a ‘threshold energy' that attracts all the initial configurations in the liquid. Our results strengthen the analogy between mean-field spin glass models and supercooled liquids.
\end{abstract}
	
\section{Introduction}
	
The long-range $p$-spin spherical model was introduced almost 30 years ago as a 
model with quenched disorder that presents the ‘same' equilibrium dynamics exhibited by the ‘simplified' mode-coupling theory (MCT) of liquid dynamics \cite{kirkpatrick_dynamics_1987,leutheusser_dynamical_1984,bengtzelius_dynamics_1984}.

Given its simple tractability, the model plays a central role in understanding the equilibrium and out-of-equilibrium phenomena of ergodicity breaking in disordered system \cite{crisanti_spherical_1992,cugliandolo_analytical_1993,franz_recipes_1995}. And it is at the core of a larger theoretical perspective, the random first order transition (RFOT), which consider the mean-field approximation as the “zeroth" order in the description of relaxation phenomena observed in real glasses \cite{biroli_random_2009}.

On another side, the p-spin spherical model is a reference model in the study of optimization algorithms. Its energy landscape is shaped by a large number of minima and some questions naturally arise. Which are the lowest minima in the energy landscape reachable starting from a random configuration? Which algorithm can reach them? 
The “best" algorithm reaching the lowest minima defines the ‘algorithmic threshold' of the model \cite{subag2018following,alaoui_algorithmic_2020}. In the p-spin spherical model, any annealing in temperature, from fast to arbitrarily slow, reaches always the same energy (called ‘threshold energy') such that an optimization seems hard if not impossible at all.

However, in a recent paper \cite{folena_memories_2019} we exhibited the emergence of a new out-of-equilibrium dynamical phase in ‘mixed' $p$-spin spherical models ---simple generalization of the ‘pure' p-spin--- in which different thermal relaxation protocols (optimization algorithms) reach different energies below the ‘threshold' one.
The considered protocols are constructed with two relaxation regimes. The system is initially equilibrated at $T_{in}$ and then quickly cooled to zero temperature.
If $T_{in}$ is chosen above and close enough to the mode-coupling temperature $\TMCT$ (temperature at which the system looses ergodicity), when cooled quickly, the system does not forget its initial condition and its final energy lies below the ‘threshold' energy. The same phenomenology is observed in simulations of structural glasses \cite{sastry_signatures_1998}. 
On the contrary, applying the same protocols to pure $p$-spin models \cite{cugliandolo_analytical_1993}, the system always reaches the ‘threshold energy', eventually forgetting any configuration $s(t')$ reached at any finite time $t'$ after the quench. 

The property of asymptotically loosing the memory of any past configuration is called weak ergodicity breaking (WEB) \cite{bouchaud_weak_1992}. 
Instead, if the system remembers the initial condition ---and any configuration $s(t')$ reached at a later time $t'$--- the relaxation dynamics has a strong ergodicity breaking (SEB) \cite{bernaschi_strong_2019}.

Pure $p$-spin models (with only one $p$-body interaction) are a special subclass of mixed $p$-spin ones (multiple $p$-body interactions). Their WEB relaxation dynamics follows from its peculiar energy landscape, where there is only one energy value (the so-called ‘threshold' energy) that attracts every out-of-equilibrium dynamics from the liquid phase. This threshold manifold is the only locus where marginal states can be found (marginality is the property of an energy minimum that makes it an attractive fixed point for the out of equilibrium dynamics at long times) \cite{cavagna_supercooled_2009}).

On the other hand, the energy landscape of mixed models is more complex, and different ‘marginal' manifolds are suitable for the asymptotic dynamics (see appendix A). As a consequence the relaxation dynamics is richer, presenting regions of both WEB and SEB, resembling what is observed in real structural glasses. This new class of models thus  enlarges the RFOT scenario in the out-of-equilibrium regime.

In order to understand this ‘richer' out-of-equilibrium dynamics, we consider the over-damped Langevin dynamics at zero temperature, i.e.\ a gradient descent dynamics in the energy landscape. The system, initially at equilibrium at $T_{in}$, relaxes at temperature $T=0$. This gradient descent dynamics will asymptotically bring the system to its inherent structures (IS), i.e.\ the configurations that correspond to local minima of the energy landscape; therefore, providing a direct connection between dynamics and geometry of the energy landscape. In the properly chosen mixed $p$-spin models (see section \ref{model}), three different dynamical-geometrical regimes are observed depending on the initial temperature $T_{in}$ \cite{folena_memories_2019}:
\begin{itemize}
\item $T_{in}>\Tonset$: the system ‘slowly' relaxes towards the ISs whose energies are temperature independent, and it forgets its initial configuration;
\item $\TSF<T_{in}<\Tonset$: the system ‘slowly' relaxes towards the ISs whose energies depend on $T_{in}$ and which lie at a finite distance from the initial configuration;
\item $\TK<T_{in}<\TSF$: the system ‘quickly' relaxes to the ISs whose energies depend on $T_{in}$ and which lie at a finite distance from the initial configuration.
\end{itemize}
In general $\TSF<\TMCT<\Tonset$ hold. While ‘slowly' indicates a relaxation where one time observables have a power law decay to their asymptotic limit, ‘quickly' refers to an exponential relaxation. The temperature $\Tonset$ was introduced in \cite{sastry_signatures_1998} to define the onset of glassiness in structural glasses. The regime above $\Tonset$ is analogous to the one found in pure $p$-spin models above $\TMCT$, in which the same ‘threshold' energy manifold attracts the dynamics from any $T_{in}$. While $\TSF$ is the state-following temperature, below which the equilibrated system is confined in a free-energy ‘well' that shrinks but doesn't break if the temperature is lowered to zero, no matter how \cite{barrat_temperature_1997}. Between $\TSF$ and $\Tonset$ rises the unexpected regime that will be the main focus of this manuscript and which is completely absent in pure $p$-spin models where $\TSF=\TMCT=\Tonset$. 

In order to test the rather unexpected ‘under-threshold' dynamics presented in \cite{folena_memories_2019} and study its finite size corrections, here we present some direct simulations of the pure $3$-spin and the mixed $(2+3)$-spin models.  Despite the importance of the model, only a few attempts to the direct simulation have been made \cite{kurchan_phase_1995,boltz_fluctuation_2019,mannelli_who_2020}. To reach large system sizes, we have considered \textit{diluted} models, with the same thermodynamic limit as the canonical one \cite{semerjian_on_the_stochastic_2004,gradenigo_solving_2020}.
Another important ingredient used in the simulation is the \textit{planting} method \cite{krzakala_hiding_2009}, which allows to prepare the system directly at equilibrium, with no need of slow annealings in temperature or other equilibration algorithms.
With these two ingredients we have simulated a $(2+3)$-spin system up to $N=2^{15}$ prepared at different temperatures $T_{in}$. 
	
As a first important result, we confirm the thermodynamic limit of the gradient descent dynamics found by the integration of mean-field dynamical equations \cite{folena_memories_2019}. In the $(2+3)$-spin spherical model, preparing the system at high temperature, $T_{in}>\Tonset$, the dynamics loses memory of the initial condition following the WEB conjecture. While preparing the system near the mode-coupling temperature $\TMCT<T_{in}<\Tonset$, the dynamics presents SEB keeping memory of the initial condition.
	
Secondly, we have simulated smaller systems of sizes $N=2000,4000$ considering a single realization of the quenched disorder and many different initial configurations at different temperatures $T_{in}$. In this case the planting method is not available, since by construction it provides just a single equlibrium configuration (on top of which the quenched disorder is build, see section \ref{dil_plant}). 
Therefore we have proceeded with a standard simulated annealing in temperature to achieve equilibrium and sample many different equilibrium configurations. Using a conjugated gradient algorithm, we find the ISs and connect their energies with the one of the initial configurations. This analysis shows that in the $(2+3)$-spin it is possible to define an onset temperature $\Tonset$, whereas by contrast, in the $3$-spin such a temperature does not exist. This confirms the importance of mixed models (rather than pure ones) in the study of complex energy landscapes.
	
The structure of the paper is the following: in section 2 the $p$-spin spherical model is introduced and the distinction between pure and mixed models is examined. In section 3 dilution and planting are defined and discussed. In section 4 the main results are reported: the agreement with the mean-field integration and the presence of the onset in finite systems. Appendix A is dedicated to the theme of marginality in mixed vs pure models, while appendix B regards the finite size scaling of ISs.

\section{Model and induced dynamics}\label{model}

The Hamiltonian of the spherical $p$-spin model is defined by a series of interaction terms with quenched disordered couplings:
\begin{equation}
	H[s] = \alpha_1 \sum_{i}J_{1}^{i}s_{i} + \alpha_2 \sum_{i,j}J_{2}^{ij}s_{i}s_{j} + \alpha_3 \sum_{i,j,k}J_{3}^{ijk}s_{i}s_{j}s_{k} + ...
\end{equation}
where spins $s_i\in\mathbb{R}$ with $i \in [1,N]$ and are confined on a sphere $\sum_{i}s_{i}^2=N$. Each term is a $p$-body interaction and it is defined by a Gaussian random tensor $J_p$ whose components are i.i.d.\ Gaussian variables with variance $\mathbb{E}[{J_p^{ij...}}^2] = \frac{1}{2} N/N^p$.
The choice of considering all possible $p$-uples instead of only the ordered ones (as in the original definition of the model \cite{crisanti_spherical_1992}) is dictated by the requirement of having a rotation-invariant system also for finite $N$, e.g.\ $P(J_{3}^{ijk}) = P(J_{3}^{i'j'k'}R_{i'}^{i}R_{j'}^{j}R_{k'}^{k})$, where $R$ is a generic rotation.
The original formulation for the Hamiltonian comes automatically contracting with the symmetric tensor of spins $s_i s_j\ldots$ and ignoring interaction terms with repeated indices which are subdominant in the large $N$ limit.

The $\alpha_p$ are parameters that define the specific model considered (hereafter we set $\alpha_1=0$ as the presence of an external field requires special care).
\textit{Pure} models are those for which only one $\alpha_p$ is different from zero and therefore they have a homogeneous Hamiltonian such that
\begin{equation}\label{homogeneous}
	\sum_{i}s_{i}\partial_{s_i} H[s] = p H[s]
\end{equation}
In this paper, we focus on the $(2+3)$-spin model which is defined by $\alpha_2=1$ and $\alpha_3=1$, while in the previous paper \cite{folena_memories_2019} we have considered the $(3+4)$-spin.
	
The variance of the Gaussian disorder Hamiltonian can be rewritten in a compact form as
\begin{equation}\label{fluctuations}
	\mathbb{E}\big[[H[s]H[\bar{s}]\big] = \sum_p \alpha^2_p \mathbb{E}[{J_{p}}^2]\Big(\sum_{i}s_{i}\bar{s}_{i}\Big)^p = N \frac{1}{2} \sum_p \alpha_p^2 q^p \equiv N f(q)
\end{equation}
where we have introduced the overlap $q=\sum_{i}s_{i}\bar{s}_{i}/N \in [-1,1]$ between spin configurations. The function $f(q)$ is the polynomial that uniquely relates to each specific $p$-spin model. By means of $f(q)$ it is possible to define different classes of models, that correspond to different kinds of ergodicity breaking at low temperature \cite{crisanti_spherical_2006}. In the following we restrict our analysis to the class of $p$-spin models that presents a phenomenology appropriate to describe structural glasses in the RFOT perspective. This RFOT-class is defined by the condition $\partial^2_q \big [ f(q)^{-\frac{1}{2}}\big ] > 0$ for every $q\in[0,1]$ and it corresponds to models having a thermodynamic transition with one step of replica symmetry breaking (1-RSB) at low temperature.
All \textit{pure} $(p>2)$-spin models have $f(q)=q^p/2$ and fall into this class.
The same is true for our reference $(2+3)$-spin model for which $f_{2+3}(q) = (q^2+q^3)/2$.
In the thermodynamic limit, these models present three phases at thermal equilibrium:
\begin{equation}
	\begin{tabular}{c c c}
		\hline 
		$\TMCT<T$ & liquid phase & $E = -f(1)/T$\\ 
		\hline 
		$\TK<T<\TMCT$ &  glassy phase & $E = -f(1)/T$\\ 
		\hline 
		$T<\TK$ & deepest glass & $E > -f(1)/T$\\ 
		\hline 
	\end{tabular}
\end{equation}
where $E \equiv \mathbb{E}[\langle H \rangle]/N$ is the average energy of the system\footnote{$\mathbb{E}$ stands for average over disorder and $\langle \rangle$ is the thermal average.}. $\TMCT$ is the mode coupling temperature that defines the breaking of ergodicity of the phase space, i.e.\ $\lim_{t \to \infty} C(t,0) = \lim_{t \to \infty}\lim_{N\to\infty} \sum_{i}s_i(t)s_i(0)/N \neq 0$ for $T<\TMCT$. At $\TK$,  the Kauzmann temperature, the Gibbs measure gets concentrated on the glassy states of lowest free-energy: in the replica formalism this corresponds to a phase transition to a 1-RSB phase. For $T>\TK$ all the equilibrium properties of the system can be evaluated by an annealed computation, which means that the fluctuations from sample to sample are so small that not only the free-energy is self-averaging, but also the partition function itself does not fluctuate in the large $N$ limit and can be computed by the following annealed average
\begin{equation}\label{annealed}
	Z = \mathbb{E}\int \mathcal{D} s\, \exp(-\beta H[s]) = \int \mathcal{D} s\, \mathbb{E}[\exp(-\beta H[s])] = \int \mathcal{D} s\, \exp\left(-\frac{\beta^2}{2} \mathbb{E}[H[s]H[s]]\right) = \exp\left(-N\frac{\beta^2}{2}f(1)\right)
\end{equation}
where $\mathcal{D} s$ is the uniform measure over the $N$-dimensional sphere defined by $\sum_i s_i^2=N$.
	
Having restricted the analysis to model belonging to the RFOT-class, we introduce the Langevin dynamics into the model \cite{crisanti_spherical_1993}:
\begin{equation}\label{langevin}
	\begin{cases}
	\partial_t s_i(t) = \sum_{k}P_i^k\big[-\partial_{s_k}H[s(t)]+\xi_k(t)\big]\\
	\langle \xi_i (t) \xi_j(t') \rangle = 2 T_{f} \; \delta_{ij}\delta (t-t')
	\end{cases}
\end{equation}
where $P_i^k=\delta_{ik}-s_is_k/N$ is the projector on the space tangent to the sphere and $\xi_i$ is the standard white noise of the thermal bath at temperature $T_{f}$.
This dynamics must be flanked with an initial condition on the spins configuration. We consider an initial condition equilibrated at inverse temperature $\beta_{in}$
\begin{equation}\label{initialEq}
	P(s(0)) \propto \exp\big(-\beta_{in} H[s(0)]\big)
\end{equation}
This defines our two-temperature protocol of dynamics: the system is prepared at the initial temperature $T_{in}$ and relaxed at the final temperature $T_f$.

This protocol has been studied in the past \cite{barrat_temperature_1997}.
There have been several attempts to understand the asymptotic solution to the out of equilibrium dynamics resulting from this two-temperatures protocol.
Unfortunately beyond some particular cases, namely equilibrium dynamics ($T_f=T_{in}$) and aging starting from a random configuration ($T_{in}=\infty$), the attempts have been inconclusive \cite{franz_recipes_1995,barrat_temperature_1997,capone_off-equilibrium_2006,sun_following_2012}.
All these attempts have used the Franz-Parisi potential \cite{franz_recipes_1995}, which is build by considering first a reference configuration $s^0$ sampled from the equilibrium distribution at $T_{in}$ and then computing the free energy cost of placing a second configuration $s^{\infty}$ (that would represents the typical configuration reached by the dynamics in the large times limit) at temperature $T_f$ and at a fixed overlap $q$ from $s^0$. 
In pure $p$-spin models this construction provides the same answer one gets from integrating the Langevin dynamics (\ref{langevin}) at very large times: that is, if $T_f=T_{in}$ in the glassy phase, the potential presents a metastable minimum at $q=\lim_{t\to\infty}C(t,0)$, and when $T_{f}$ is lowered this metastable minimum keeps corresponding to the long time dynamics. This procedure is called state-following in temperature but unfortunately does not work for mixed $p$-spin models since the metastable minimum in this class of models disappears for $T_{f}<T_{lost}(T_{in})$ and there is no alternative recipe for computing the large time asymptotic of the two-temperatures protocol\footnote{Such a clear distinction between pure and mixed $p$-spin models has been recently confirmed also via the study of the minima of the Thouless-Anderson-Palmer free energy \cite{barbier_constrained_2020}.}.
The lack of any analytic tool to predict the large times dynamics in mixed $p$-spin models makes the integration of the dynamical equations the only way to understand the outcome of the two-temperature protocol.

Focusing on the case of $T_f=0$, which corresponds to a gradient descent in the energy landscape, we have found the following scenario \cite{folena_memories_2019}: 
\begin{equation}
	\begin{tabular}{c c c c}
		\hline 
		\textit{pure} RFOT models & $T_{in}>\TMCT$ aging with WEB & $E = E_{th}$ & $\mu = \mu_{mg}$\\
		& $T_{in}<\TMCT$ relaxation towards IS of the glass & $E < E_{th}$ & $\mu > \mu_{mg}$\\
		\hline 
		\textit{mixed} RFOT  models & $T_{in} \gtrsim \Tonset$ aging with WEB & $E = E_{rc}$ & $\mu = \mu_{mg}$\\ 
		& $\TSF<T_{in} \lesssim \Tonset$ aging with SEB & $E < E_{rc}$ & $\mu = \mu_{mg}$\\ 
		& $T_{in}<\TSF$ relaxation towards IS of the glass & $E < E_{rc}$ & $\mu > \mu_{mg}$\\
		\hline 
	\end{tabular}
\end{equation}
where $\mu=-\sum_i s_i \partial_{s_i}H[s]/N$ is the radial reaction, i.e.\ the radial force exerted at $s_i$ by the spherical constraint to maintain the spin configuration on the sphere. As shown in appendix A, if $s$ is a stationary point of the energy $H[s]$, then $\mu$ controls the value of the lowest eigenvalue associated to its Hessian, $\mathcal{H}_{ij}=\partial^2 H[s]/\partial_{s_i}\partial_{s_j}$.
If $\mu>\mu_{mg}$ the stationary point is a minimum, while for $\mu<\mu_{mg}$ it is a saddle. The ‘marginal' minima are defined by the condition $\mu=\mu_{mg}$ and it is important to remind that the Langevin dynamics at large times converges to marginal minima with high probability.
 	
In pure $p$-spin models, all marginal states have a well-defined energy, called threshold energy $E_{th}=-\mu_{mg}/p$, and the Langevin dynamics starting from any temperature in the liquid phase ($T_{in}> \TMCT$) converges to $E_{th}$. At variance, in a mixed $p$-spin model there is a entire range of energies where marginal states can be found \cite{folena_memories_2019} and predicting which one is going to attract the out-of-equilibrium Langevin dynamics is highly non trivial. Starting from a random configuration the dynamics converges (with high probability in the large $N$ limit) to the energy $E_{rc}$ whose value is reported in Eq.~(\ref{threshold}) in appendix A (where more details on the differences between pure and mixed models are also provided). Starting from $T_{in}\lesssim \Tonset$ the Langevin dynamics goes ``under the threshold'' and reaches energies lower than $E_{rc}$, while keeping memory of the initial configuration, i.e.\ showing SEB.
The $\Tonset$ temperature was introduced in \cite{sastry_signatures_1998} in order to describe the dependence of the inherent structure (IS) energy on the temperature $T_{in}$.
Lowering further $T_{in}$ below $\TSF$ the aging ceased and the Langevin dynamics relaxes within a state, that can be followed down to $T_f=0$, since $T_{lost}(\TSF)=0$ \cite{folena_memories_2019}.
In between the two temperatures $\TSF$ and $\Tonset$ a new out-of-equilibrium phase emerges, which shows aging behavior in a restricted part of the phase space. 
Its full understanding is far to be complete.
	
To conclude the section we report some relevant temperatures in the mixed ($2+3$)-spin model and in the pure $3$-spin model:
\begin{center}
\begin{equation}
	(2+3)\text{-spin}: 
	\begin{tabular}{|c|c|c|c|c|c|}
		\hline 
		$\TSF$ & $\TK$ & $\TMCT$ & $\Tonset$  & $\mu_{mg}$ & $E_{rc}$\\ 
		\hline 
		$<\TK$ & 1.0185.. & 1.0206.. & 1.25 & 4 & -1.55\\	\hline 
	\end{tabular}
\end{equation}	
\begin{equation}
\vspace{0.5cm}
	3\text{-spin}: 
	\begin{tabular}{|c|c|c|c|}
		\hline 
		$\TK$ & $\TMCT$ & $\mu_{mg}$ & $E_{th}$\\ 
		\hline 
		0.5861.. & 0.6124.. & -3$E_{th}$ & -1.1547..\\ 
		\hline 
	\end{tabular}
\end{equation}			
\end{center}
We observe that in the $(2+3)$-spin, contrary to the $(3+4)$-spin studied in Ref.~\cite{folena_memories_2019}, $\TSF$ is below $\TK$ and so no equilibrium glass prepared above $\TK$ can be followed down to $T_f=0$.
Moreover, we notice that $\TK$ and $\TMCT$ are very close to each other since number of metastable states in this model is much smaller than in the $(3+4)$-spin studied in Ref.~\cite{folena_memories_2019}.
For a thorough discussion of equilibrium and out-of-equilibrium aspects of the model we refer the reader to \cite{folena_mixed_2020}.

\section{Simulation Preliminaries: Dilution and Planting}\label{dil_plant}
	
In order to substantiate the results obtained in Ref.~\cite{folena_memories_2019} in the $N\to\infty$ limit, we have performed a numerical simulation of the gradient descent dynamics ($T_f=0$) for a mixed $p$-spin model in the RFOT class, starting from different initial temperatures $T_{in}$. Instead of the $(3+4)$-spin model studied in the aforementioned paper, we have considered the $(2+3)$-spin model because of numerical convenience, given it has $O(N^3)$ interactions rather than $O(N^4)$.

To reach sizes such that the finite size effects on a single trajectory are tiny ($N\gtrsim2^{12}$) we resort to the dilution of the interaction terms, preserving the thermodynamic limit \cite{krzakala_performance_2013}. 
We select a random fraction $D_p$ of all interactions $J_p$ and since we want to have the same statistics in the thermodynamic limit, i.e.\ the same energy correlations as in Eq.~(\ref{fluctuations}), we rescale accordingly the variance of the quenched disorder to $\mathbb{E}[{J_p}^2] = \frac{1}{2} N/(N^p D_p)$. As a result we obtain a diluted Hamiltonian with less interaction terms, but each one has a larger strength. However, there is a caveat: the dilution parameter $D_p$ cannot be arbitrarily small, otherwise the system will \textit{condensate}. This may happens since, upon strong dilution, the system becomes more and more heterogeneous to the point that the minimum energy is achieved on configurations where a vanishing fraction (in the large $N$ limit) of spins becomes exceedingly large, while the majority of spins take values close to zero \cite{semerjian_on_the_stochastic_2004,gradenigo_solving_2020}.
To compute a lower bound to the dilutions that avoid condensation we consider the worst-case scenario where a single spin gets the whole weight\footnote{This argument it is usually done considering a subset of $p$ spins, instead of a single spin, that get the whole weight, but in our model definition there are diagonal terms (that were ignored in previous studies). In any case the two arguments provide the same scaling.}.
Setting $s_{i_{\rm min}}=\sqrt{N}$ with $i_{\rm min} = {\rm argmin}_i[J_p^{ii..i}]$ and $s_i=0$ for $i \neq i_{\rm min}$ we get the energy of the most condensed configuration
\begin{equation}
N E_{cond} = J_{\min} s_{i_{\min}}^p =  -\sqrt{2 \log N} \sqrt{\mathbb{E}[{J_p}^2] N^{p}} 
	=  -\sqrt{2 \log N} \sqrt{\frac{1}{2} N / D_p}
\end{equation}
where $J_{\rm min}=\min_i[J_{p}^{ii..i}]$ and the term $\sqrt{2 \log N}$ comes from the largest fluctuations of extreme values of Gaussian distributed random variables \cite{gumbel2012statistics}. This result is identical for every $p$-body interaction. $E_{cond}$ should be compared with the typical energy $E_{sim}$ that we need to study in our simulation. 
\begin{equation}
	E_{sim} < E_{cond} \quad \Longrightarrow \quad D_p > \frac{\log N}{N E^2_{sim}}
	\label{lowerDilution}
\end{equation}
For the gradient descent dynamics in the $(2+3)$-spin we use $E_{sim} = E_{3,rc} \approx -0.9$, i.e.\ the $3$-body part of energy reached starting from a random initial condition\footnote{$E_{rc}=E_{2,rc}+E_{3,rc}=-1.55$}. This allows to avoid condensation during the dynamics. In case one needs to further increment the dilution, the simplest way is to take a bounded distribution of interactions. This erases the extreme value factor $2 \log N$, giving the relation $D_p > 1/(2 N E^2_{sim})$. 
	
In simulating the $(2+3)$-spin model we have chosen different dilutions, both with Gaussian and bimodal distribution of the couplings, trying to stay close to the lower bound in Eq.~(\ref{lowerDilution}).
Close, but definitely above it, usually by a factor around 2.
In this way we have been able to simulate systems up to $N=2^{16}$.

Having prepared the quenched interaction we need to extract an initial configuration at equilibrium at $T_{in}$. This can be achieved with an annealing in temperature till the desired temperature. For large systems, the equilibration time near $\TMCT$ is very large and equilibration becomes computationally expensive. So we resort to a trick which is always available in mean-field models whenever the typical fluctuations induced by the quenched disorder are small and the annealed average is correct, i.e.\ if $\mathbb{E}[\log(Z)] = \log(\mathbb{E}[Z])$. This is the case for $T_{in}>\TK$ as we have already discused around Eq.~(\ref{annealed}). In this region it is possible to \textit{plant} the initial configuration following Ref.~\cite{krzakala_hiding_2009}. First a spin configuration $s^*$ is randomly chosen on the sphere and then the couplings $J_p$ are extracted according to a $s^*$-tilted Gibbs distribution. Thus, we have a new Hamiltonian $H^*[s]$ such that $s^*$ is an equilibrium configuration at $T_{in}=1/\beta_{in}$:
\begin{equation}
	H^*[s] = H[s]-\frac{N \beta_{in}}{2}\left(\alpha^2_1 \sum_{i}\frac{s^*_{i}}{N}s_{i} + \alpha^2_2  \sum_{ij}\frac{s^*_{i}s^*_{j}}{N^2}s_{i}s_{j} + \alpha^2_3 \sum_{ijk}\frac{s^*_{i}s^*_{j}s^*_{k}}{N^3}s_{i}s_{j}s_{k} + \ldots\right)
\end{equation}
It is easy to show that $\mathbb{E}[H^*[s^*]] = -N \beta_{in} f(1)$.
Therefore the new interactions are just shifted by the respective tensor of order $p$ built upon $s^*$, which for our reference ($2+3$)-spin gives
\begin{equation}
{J^*_2}^{ij} = J_2^{ij}-\frac{\beta_{in}}{2}\frac{s^*_{i}s^*_{j}}{N^2} \qquad {J^*_3}^{ijk} = J_3^{ijk}-\frac{\beta_{in}}{2}\frac{s^*_{i}s^*_{j}s^*_{k}}{N^3}
\end{equation}
where $J_2^{ij}$ and $J_3^{ijk}$ are extracted from the original Gaussian or bimodal distribution.

\section{Gradient Descent Simulation}
	
In this section we report the main results. Firstly we show the agreement between the simulation of the gradient descent dynamics in large systems and the integration of the MFDE.
We show the results for two system sizes, $N=2^{13}$ and $N=2^{15}$, using planting together with dilution of the interactions as described in the previous section. 
	
In the second part of this section we study the energy landscape for a single realization of the quenched disorder. We consider two systems of sizes $N=2000$ and $N=4000$ both in pure and mixed models and study how the ISs depend on $T_{in}$. Already at these small sizes, mixed models present an onset temperature and the related dynamics going below $E_{rc}$, while pure models presents a unique ‘threshold energy'.
	
\subsection{Agreement with Mean-field Integration}
		
Having introduced the two-temperature ($T_{in},T_{f}$) protocol, given by the Langevin dynamics in Eq.~(\ref{langevin}) together with the starting equilibrium condition in Eq.~(\ref{initialEq}), it is possible, through two different approaches ---dynamical cavity approach and path integral formalism--- to derive the correspondent mean-field dynamical equations (MFDE). These equations relate the two-time correlation $C_{tt'} \equiv \sum_{i}\langle s_i(t)s_i(t')\rangle/N$ with the two-time response $R_{tt'} = \sum_{i} \partial_{h_i(t')} \langle s_i(t)\rangle/N$ of the system in the thermodynamic limit ($N\to\infty$).
\begin{equation}\label{mfde}
    \begin{aligned}
	\partial_t C_{tt'} =& -\mu_t C_{tt'}+\int_{t'}^{t} f''(C_{ts})R_{ts}C_{st'}ds
	+\int_{0}^{t'} \big ( f''(C_{ts})R_{ts}C_{t's} +f'(C_{ts})R_{t's} \big )ds+\beta_{in} f'(C_{t0})C_{t'0}\\
	\partial_t R_{tt'} =& \;\delta_{tt'}-\mu_t R_{tt'}+\int_{t'}^{t} f''(C_{ts})R_{ts}R_{st'}ds\\
	\end{aligned}
\end{equation}
where $\mu_t \equiv  T_{f}+\int_{0}^{t}  \big ( f''(C_{ts})R_{ts}C_{ts}ds + f'(C_{ts})R_{ts} \big ) ds +\beta_{in}f'(C_{t0})C_{t0}$ in order to enforce the spherical constraint. $\delta_{tt'}$ is the Dirac delta. These equations have a hidden arbitrary time scale, but for simplicity we have chosen to fix it with the normalization $\lim_{t \rightarrow t'}\partial_t C_{tt'} =-T_{f}$.
The average energy of the system is:
\begin{equation}\label{defE}
    E(t) \equiv \lim_{N\to\infty}\mathbb{E}[\langle  H[s(t)] \rangle]/N =-\int_{0}^{t} f'(C_{ts})R_{ts}ds -\beta_{in} f(C_{t0})
\end{equation}
For $T_{f}=0$, i.e.\ considering gradient descent dynamics, the equations in Eq.~(\ref{mfde}) can be integrated numerically till times of the order of $10^3$. The algorithm we have used is a simple fixed-step integration and it is reported in \cite{franz_mean_1994}. 
The plane $(t,t')$ is discretized in both dimensions by time steps of fixed size $\Delta t$ and the integration is performed on this grid, thus the total computing time grows as $(t/\Delta t)^3 = ({\rm \#steps})^3$. 
The true values at each time are recovered by extrapolating $\Delta t\to 0$\footnote{three different steps are considered and then a quadratic fit is performed}.

\begin{figure}[t]
	\centering
	\begin{minipage}[b]{0.38\textwidth}\centering
	    \includegraphics[width=0.99\columnwidth]{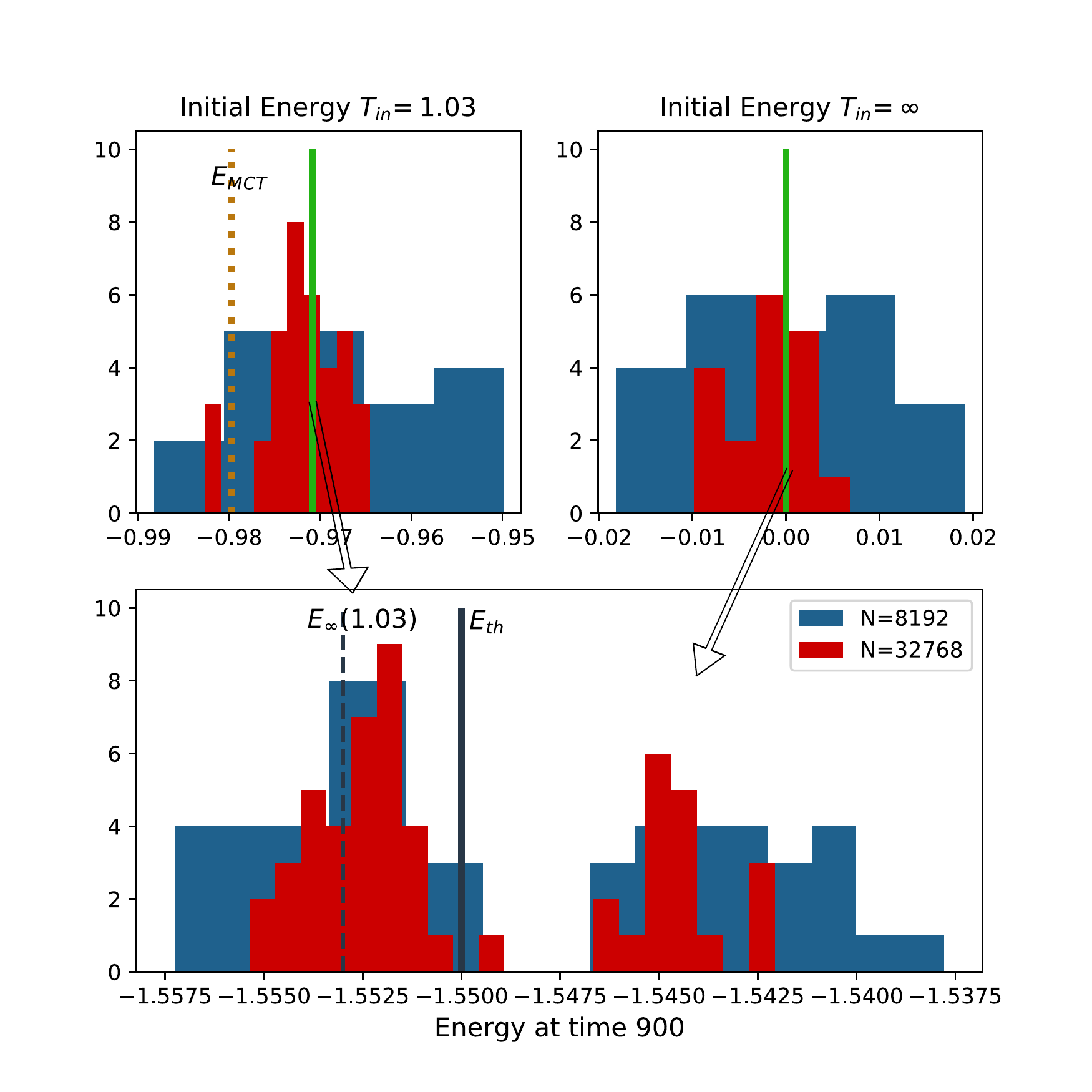}
	\end{minipage}
	\hfill
	\begin{minipage}[b]{0.6\textwidth}\centering
		\includegraphics[width=0.99\columnwidth]{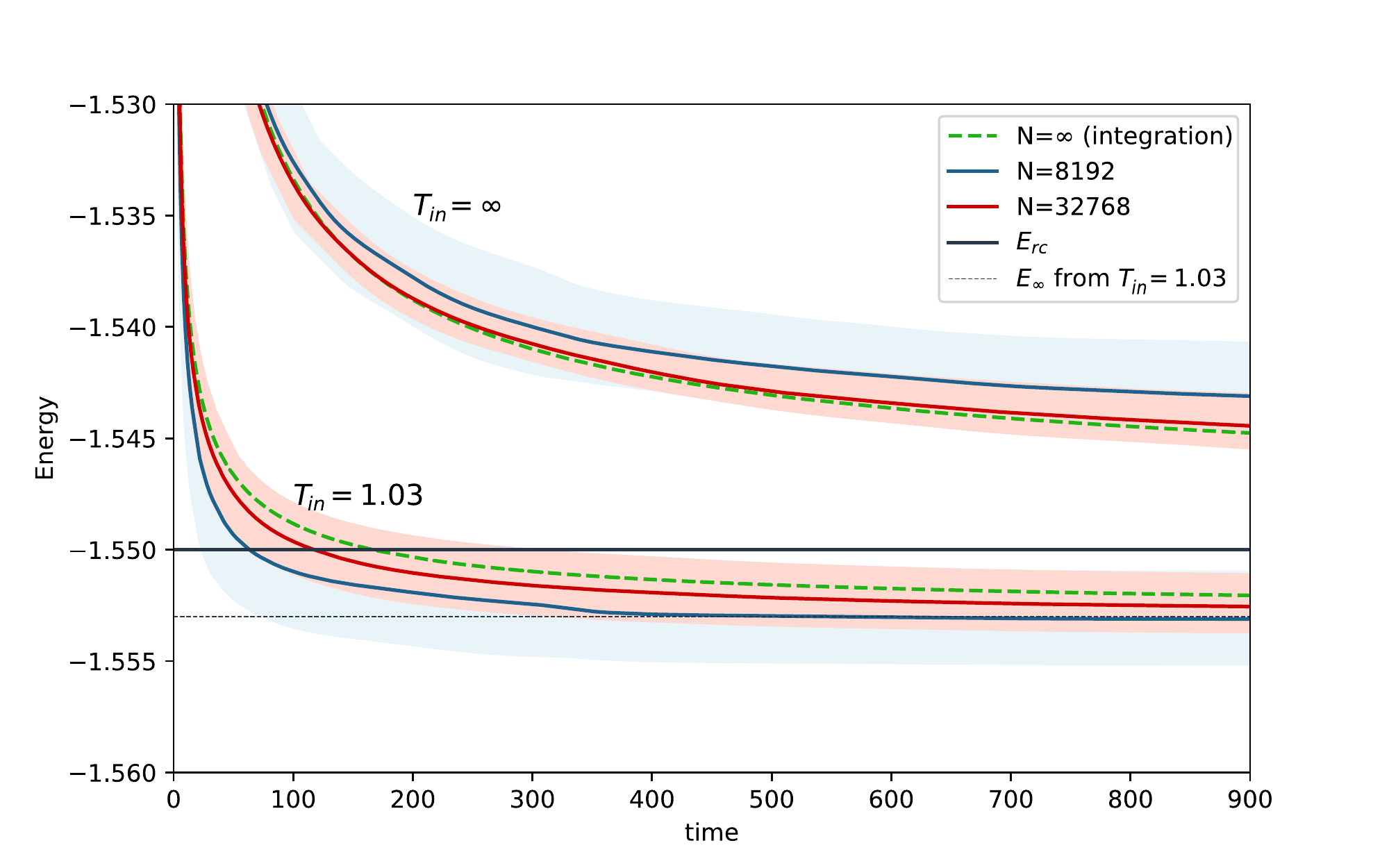}\\
	\end{minipage}
	\caption{Gradient descent dynamics in the $(2+3)$-spin model with $N=2^{13}$ and $N=2^{15}$ from random initial condition ($T_{in}=\infty$) and from equilibrium near $\TMCT$ ($T_{in}=1.03$). \textbf{Left}: above the histograms of initial energies; below the energies for the same samples at time 900. \textbf{Right}: average over the different samples of the energy as a function of time. The $N\to\infty$ line is given by the numerical integration of MFDE. The shadowed areas show the standard deviation over different trajectories. Each trajectory has a different quenched disorder.}
	\label{fig:GD1}
\end{figure}
	
\begin{figure}[t]
	\centering
	\begin{minipage}[b]{0.38\textwidth}\centering
		\includegraphics[width=0.99\columnwidth]{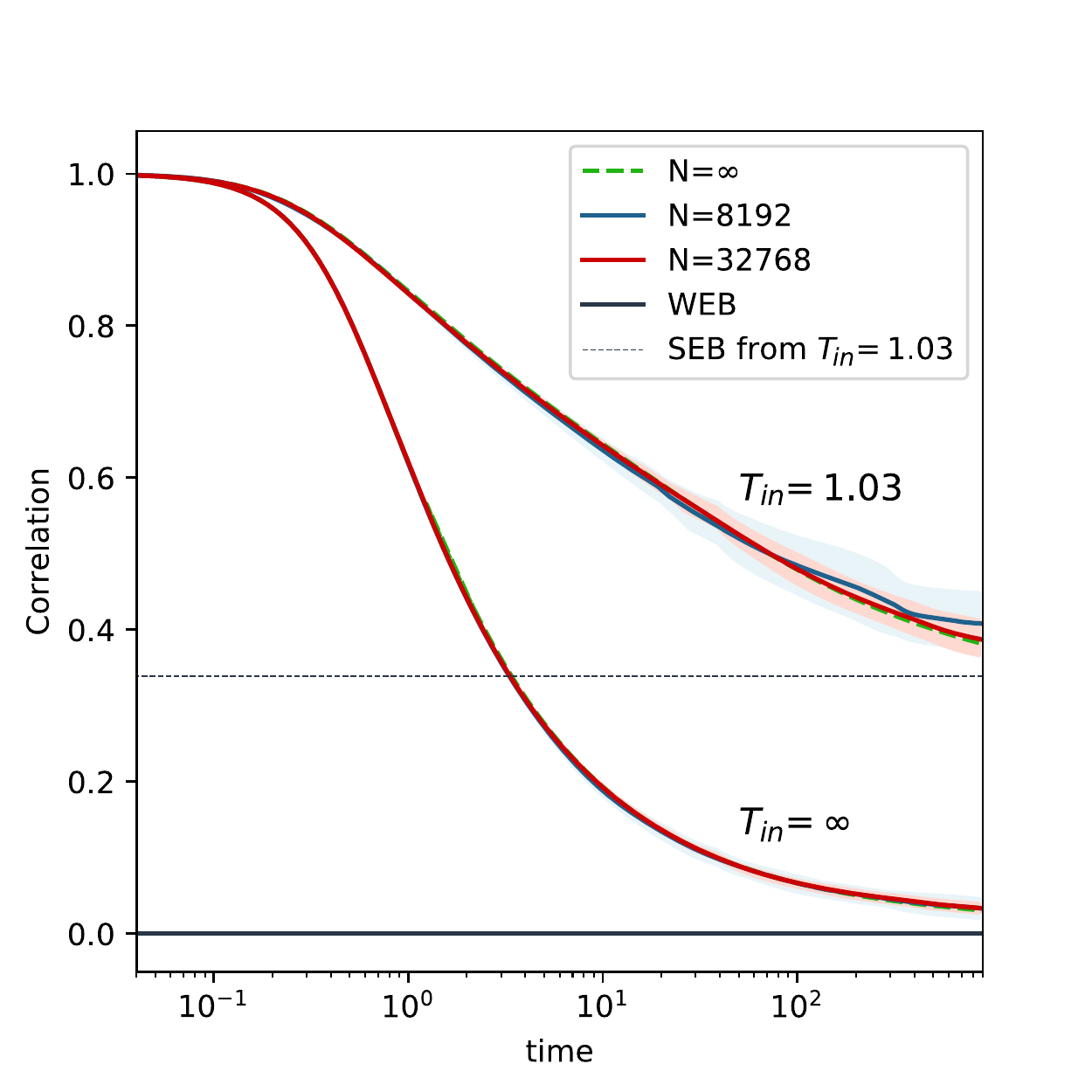}
	\end{minipage}
	\hfill
	\begin{minipage}[b]{0.6\textwidth}\centering
		\includegraphics[width=0.99\columnwidth]{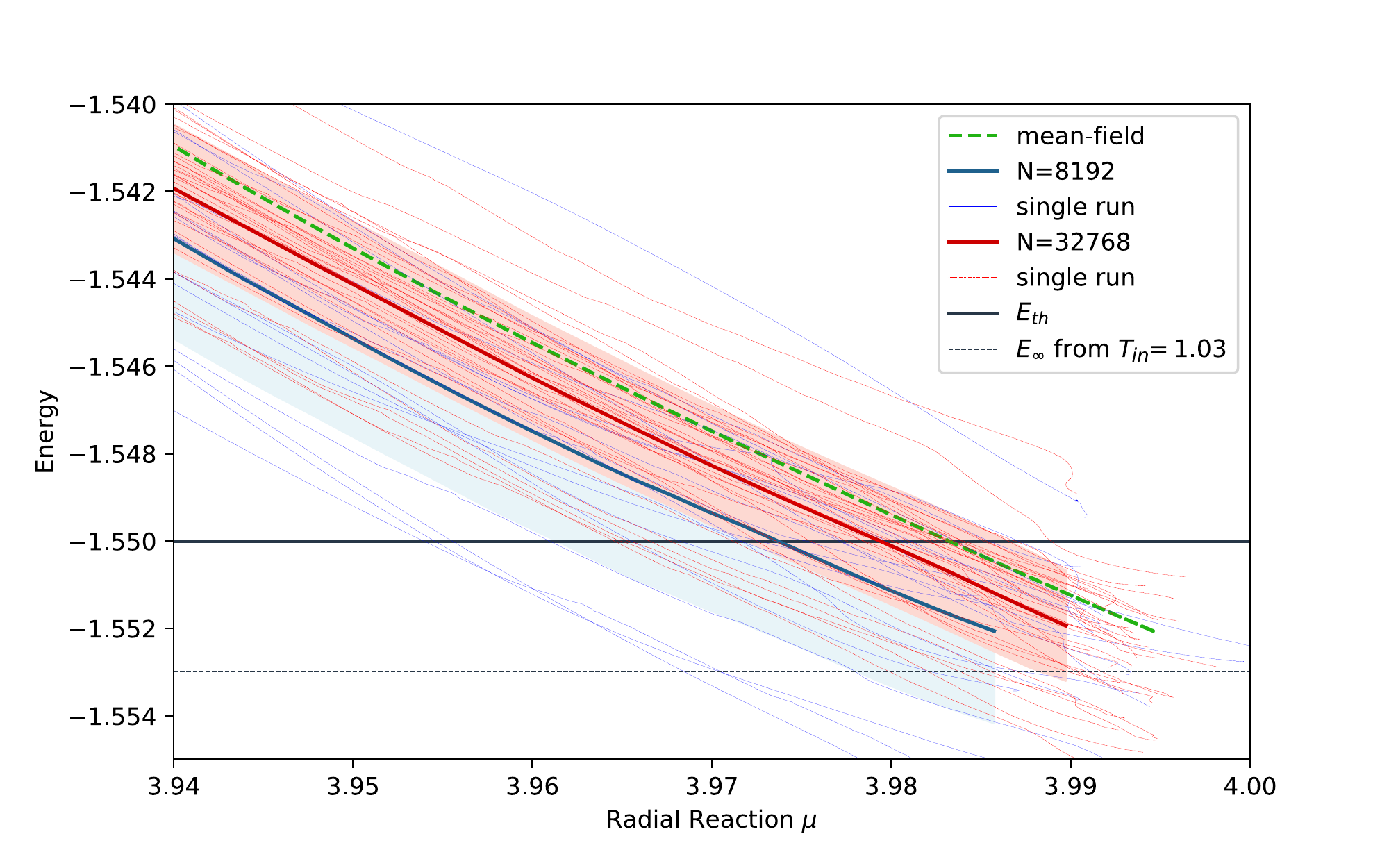}\\
	\end{minipage}
	\caption{The same samples considered in Fig.~\ref{fig:GD1}. \textbf{Left}: Correlation with the initial configuration vs time. Starting from a random configuration the memory of the initial condition is lost (WEB), while starting from $T_{in} =1.03\approx \TMCT$ strong ergodicity breaking (SEB) is observed. The dashed gray lines is an extrapolation of the mean-field  integration for infinite time. \textbf{Right}: the energy during gradient descent from $T_{in} =1.03$ as a function of the radial reaction $\mu$ until time 900. Thick lines represent the averages over all samples, while thin lines are single samples.}
	\label{fig:GD2}
\end{figure}

In Fig.~\ref{fig:GD1}, for two different system sizes, $N=2^{13}$ (blue) and $N=2^{15}$ (red), we consider the gradient descent dynamics, starting both from random configuration ($T_{in} =\infty$) and from equilibrium near $\TMCT$ ($T_{in} =1.03$). The second condition is implemented through the planting procedure, thus each trajectory corresponds to a different quenched disorder.
Here we have considered Gaussian interactions with dilutions $D_2=0.1$ and $D_3=0.001$ for systems $N=2^{13}$ and $D_2=0.025$ and $D_3=0.00025$ for systems of size $N=2^{15}$.

In the upper part of Fig.~\ref{fig:GD1} (left) we plot the histograms of the initial equilibrium energies of the system, which stay around the thermodynamic limit (green lines). In both cases fluctuations are of order $O(N^{-1/2})$, as expected. The gradient descent dynamics is implemented starting from the planted initial configuration $s(0)=s^*$. In order to enforce the spherical constraint the spin $s$ is rotated at each step in the direction defined by the projected gradient $\nabla^P H[s] \equiv P[\nabla H[s]]$ as in Eq.~(\ref{langevin}). To integrate the gradient descent dynamics we use a variable time step $\Delta t$, such that the angle of rotation $\theta$ between two consecutive projected gradients\footnote{$\sum_{i} \nabla^P_{i}H[s(t)] \nabla^P_{i}H[s(t+\Delta t)] \approx \cos(\theta) \|\nabla^PH[s(t)]\| \|\nabla^PH[s(t+\Delta t)]\|$.} is kept roughly constant, $\theta\approx0.025$.
In the limit $\theta\to 0$ the dynamics is equivalent to the Langevin dynamics at $T_f=0$.

In Fig.~\ref{fig:GD1} (right) we show the average (full line) and the standard deviation (light shadow) of the energy $E(t)$ for the two initial conditions. The green dashed line reports the results from the numerical integration of MFDE in Eq.~(\ref{mfde}). The black thick line is the $N\to\infty$ threshold energy as defined in Eq.~(\ref{threshold}). The black dotted thin line represents the long time asymptotic extrapolation of the energy starting at $T_{in}=1.03$. The dynamics are shown till time 900 and the relative histogram of energies reached at that time is shown in the bottom of Fig.~\ref{fig:GD1} (left).
Let us notice that, based on finite size considerations (appendix B) we would expect smaller systems to reach higher energies, independently of the starting temperature. However in Fig.~\ref{fig:GD1} (right) we see that for $T_{in}=1.03$ the smaller system reaches smaller energies. This is an artifact of the planting procedure in finite size systems, which is not observed when the system is prepared through annealing. In fact, for $N=8192$ some samples prepared at $T_{in}=1.03$ have a starting energy below $\EMCT$, therefore being inside a glassy state, that goes to a very low energy when cooled down\footnote{A more detailed discussion on the two procedures to equilibrate a system (planting vs.\ annealing) can be found in section 3.3.2 of Ref.~\cite{folena_mixed_2020}.}.

In Fig.~\ref{fig:GD2} (left) the average correlation with the initial configuration $C_{t0}=\sum_{i}s_i(t)s_i(0)/N$ and its standard deviation are shown. Starting from random condition WEB holds, while starting from $T_{in} =1.03\approx \TMCT$ the systems is confined in a partition of the phase space and SEB holds. The black dotted thin line marks the long-time extrapolation from the integration of MFDE. Clearly the fact that SEB really holds asymptotically in the thermodynamic limit cannot be confirmed by these simulations. Here we just want to support the correctness of the integration. To conclude in Fig.\ref{fig:GD2} (right) we show the parametric plot of the energy vs the radial reaction starting from $T_{in}=1.03$. In the thermodynamic limit the marginal radial reaction is $\mu_{mg}=4$ which is reached asymptotically in the long time limit. 
	
\subsection{Onset in Finite Systems}\label{onset}

In this section we show that the presence of an onset temperature can be observed even when simulating a single realization of the quenched disorder. We consider one system of size $N=2000$ and another of size $N=4000$ and for each one we take many initial conditions at different temperatures.
We consider smaller sizes driven by the necessity of equilibrating the system through an annealing protocol for every $T_{in}$. This is a direct consequence of the limit of the planting procedure which does not allow to simulate more than one temperature for the same disorder, since the disorder is built according to the planted configuration. The dilution considered is $D_{2}=3/\sqrt{N},D_{3}=3/N$ with bimodal interactions. In order to select the initial conditions at $T_{in}$, we have adopted a simple Monte Carlo annealing starting from $T=2$ and lowering the temperature with a constant rate $\Delta T=-0.0001$ per MC step, followed by an evolution at the constant temperature $T_{in}$ for a time long enough to allow the recollection of several independent initial configurations. In practice we require the overlap between consecutive selected configurations to be 0.1 or smaller. From each of these configurations, a gradient descent  dynamics is performed in order to end up in the correspondent IS. In sampling initial configurations we consider the same total time for every $T_{in}$. As a consequence, the total number of samples is inversely proportional to the relaxation time at that temperature, and for $T_{in}$ approaching $\TMCT$ we have very few samples.
	
\begin{figure}[t]
	\centering
	\begin{minipage}[b]{0.7\textwidth}\centering
	    \includegraphics[width=0.99\columnwidth]{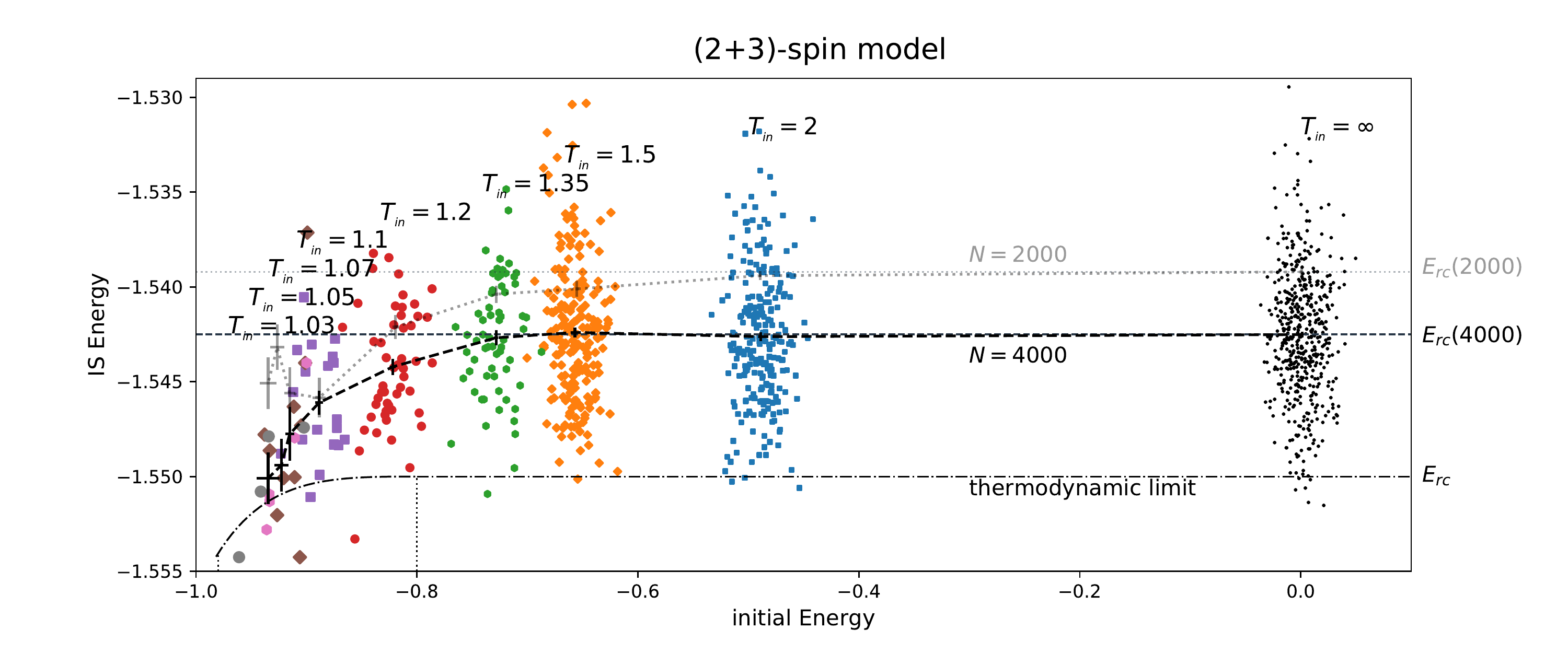}
	\end{minipage}
	\begin{minipage}[b]{0.29\textwidth}\centering
		\includegraphics[width=0.99\columnwidth]{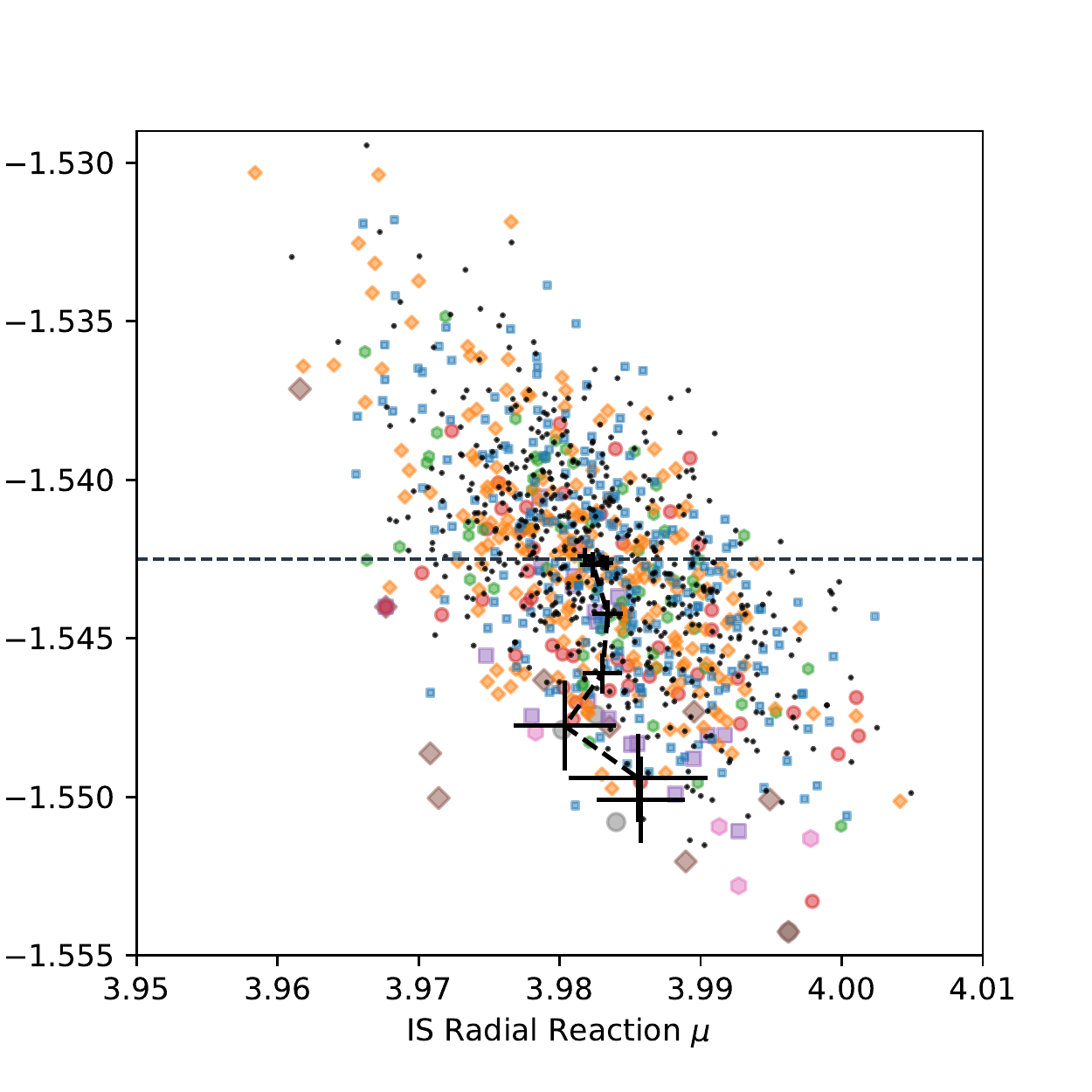}\\
	\end{minipage}
	\caption{\textbf{Left}: The energies of the IS vs those of the initial configurations for different temperatures $T_{in}$ (different colors) in a single sample of size $N=4000$ of the ($2+3$)-spin model. The dashed thick black line joins the average values computed for each temperature, while the thin dashed horizontal line marks the empirical threshold energy defined by the average energy of the IS reached from random initial conditions.  For $T_{in} \lesssim \Tonset \approx 1.35$ the dynamics goes clearly below the threshold. The dashed-dotted line corresponds to the thermodynamic limit.
	The light gray dashed curves are for the $N=2000$ sample, that seems to have a slightly higher $\Tonset$. \textbf{Right}: The same points are presented in the plane IS energy vs IS radial reaction, together with their averages (one big cross per temperature). From these averages we notice that the mean energy decreases, while the mean radial reaction remains roughly constant while lowering $T_{in}$.}  
	\label{fig:23spin}
\end{figure}
	
\begin{figure}[t]
	\centering
	\begin{minipage}[b]{0.7\textwidth}\centering
		\includegraphics[width=0.99\columnwidth]{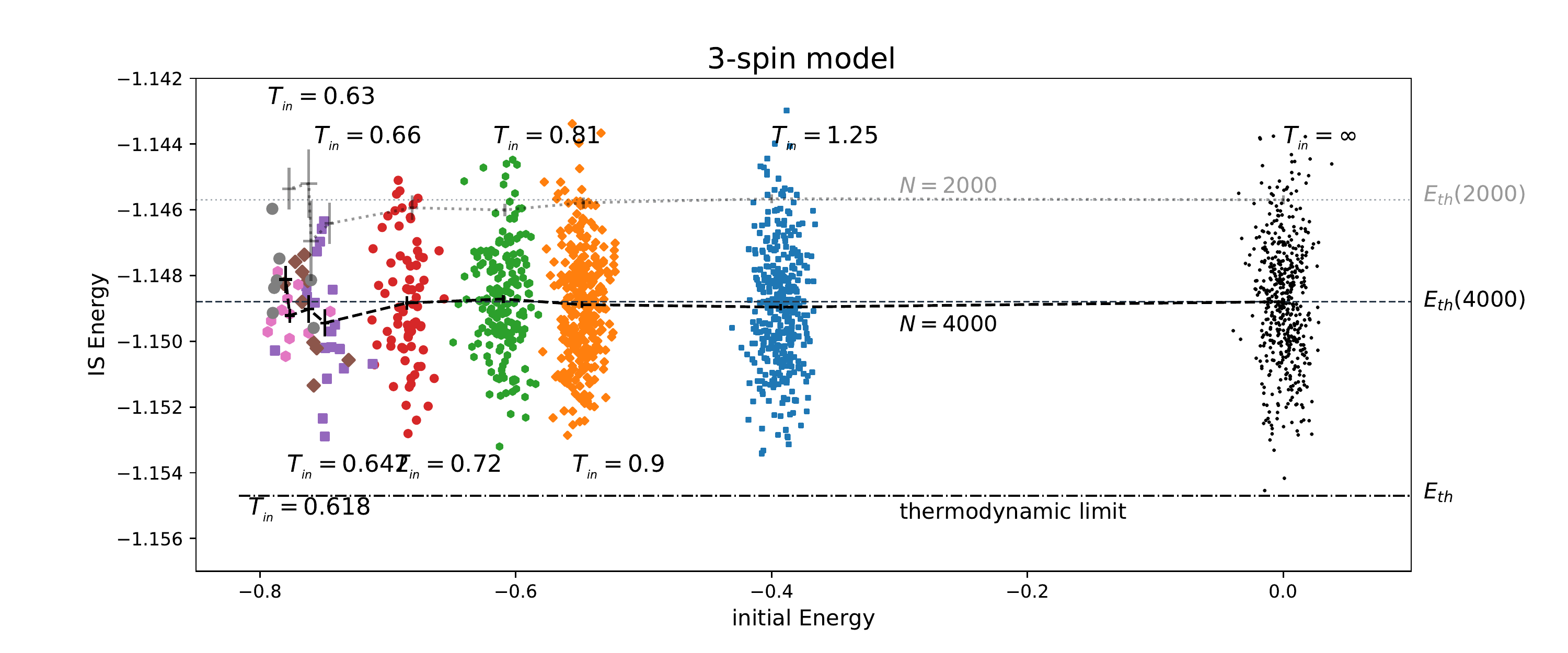}
	\end{minipage}
	\begin{minipage}[b]{0.29\textwidth}\centering
		\includegraphics[width=0.99\columnwidth]{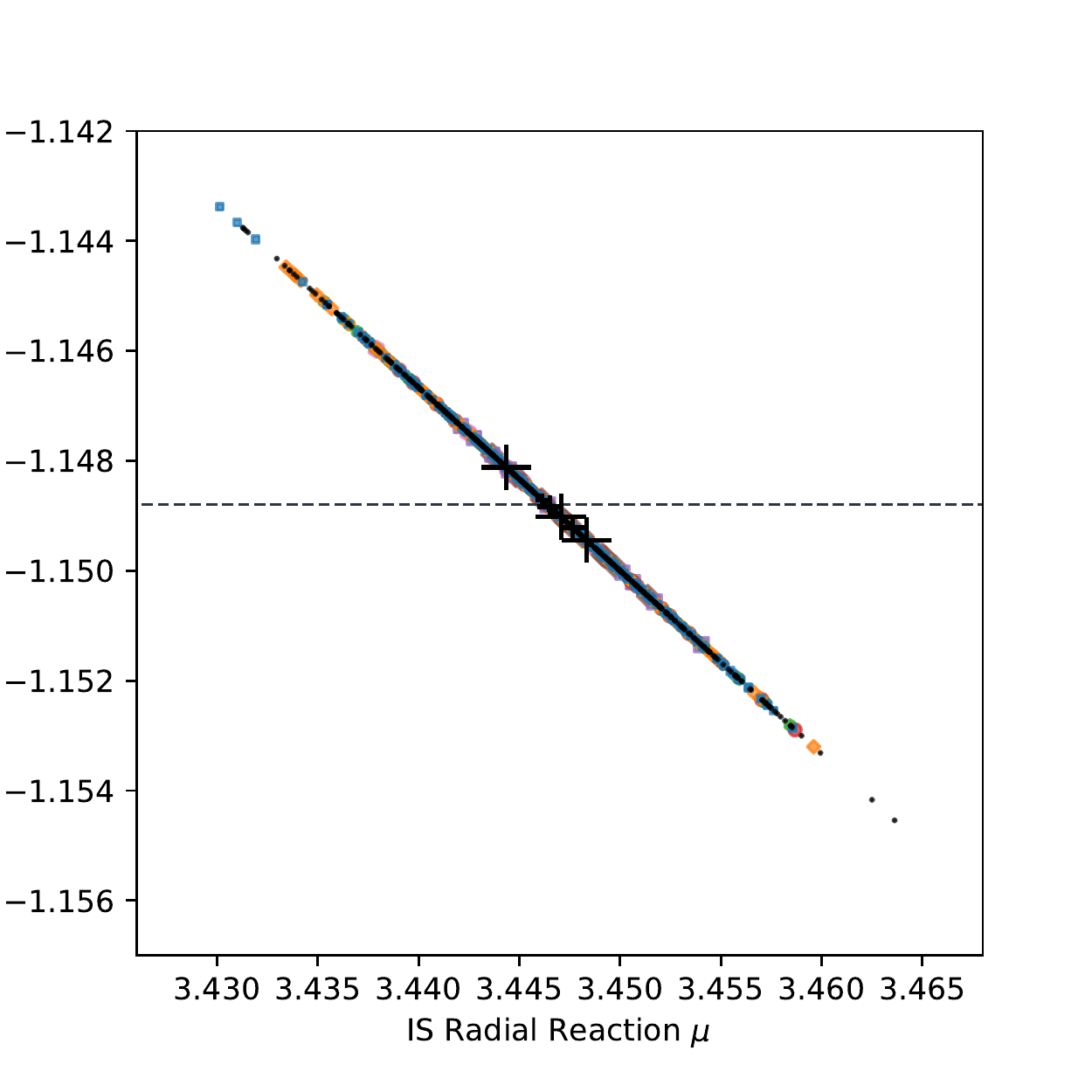}\\
	\end{minipage}
	\caption{The same analysis presented for the ($2+3$)-spin model in Fig.~\ref{fig:23spin} is here repeated for the $3$-spin model. In this case different initial temperatures $T_{in}$ have the same IS energy (\textbf{left}), since the homogeneity of the Hamiltonian implies that energy and radial reaction are proportional (\textbf{right}).}
	\label{fig:3spin}
\end{figure}

This simulation is performed both on a mixed $(2+3$)-spin and a pure $3$-spin model. 
In Fig.~\ref{fig:23spin} we show the energy of the IS as a function of the energy of the corresponding initial equilibrium configuration, for different temperatures $T_{in}$ (different colors) in the ($2+3$)-spin model. We define the random condition energy $E_{rc}$ of each system as the average IS energy reached from random initial condition (black points). Preparing the $N=4000$ system with $T_{in} \lesssim \Tonset \approx 1.35$, the IS energy average goes below the defined $E_{rc}$, therefore we can define a crossover at $\Tonset$ which in the thermodynamic limit is expected to reach $\Tonset = 1.25$. This onset temperature marks a crossover, as transition temperatures in finite systems do, and it exhibits a finite size dependence, shifting up for smaller system sizes (see the light gray dotted line for $N=2000$).

In the right plot of Fig.~\ref{fig:23spin} we report for the same ISs of the $N=4000$ system the energy $E_{\IS}$ vs the radial reaction $\mu_{\IS}$.  The black crosses (and the dashed line connecting them) represent the average and relative errors for different temperatures $T_{in}$. While $E_{\IS}$ decreases with the temperature for $T_{in} < \Tonset$, the radial reaction $\mu_{\IS}$ is roughly temperature independent, which seems to confirm that radial reaction is related to the stability of the minima reached by the relaxation dynamics also in finite size systems. In appendix B we report a finite size analysis of $E_{\IS}$ and $\mu_{\IS}$.

The same analysis is repeated for the $3$-spin in Fig.~\ref{fig:3spin}. In this case the system, prepared at different $T_{in}>\TMCT$, always relaxes towards the same threshold energy $E_{th}$ defined from random initial conditions. In this case, any annealing preparation of the system does not have any benefit on average.  This behavior is due to the homogeneity of the Hamiltonian which implies that IS minima satisfy the relation $E_{\IS}=-3 \mu_{\IS}$ (see appendix A). Hence, fixing marginality $\mu_{\IS}=\mu_{mg}$ implies fixing the energy $E_{\IS}=E_{th}=-3 \mu_{mg}$ and fluctuations of the second are implied by fluctuations of the first, which is related to fluctuations of the lowest eigenvalue of the Hessian computed at the IS \cite{boltz_fluctuation_2019}. The proportionality between $\mu_{\IS}$ and $E_{\IS}$ in the 3-spin model is highlighted by the right plot in Fig.~\ref{fig:3spin}.

\begin{figure}[t]
	\centering
	\includegraphics[width=0.9\columnwidth]{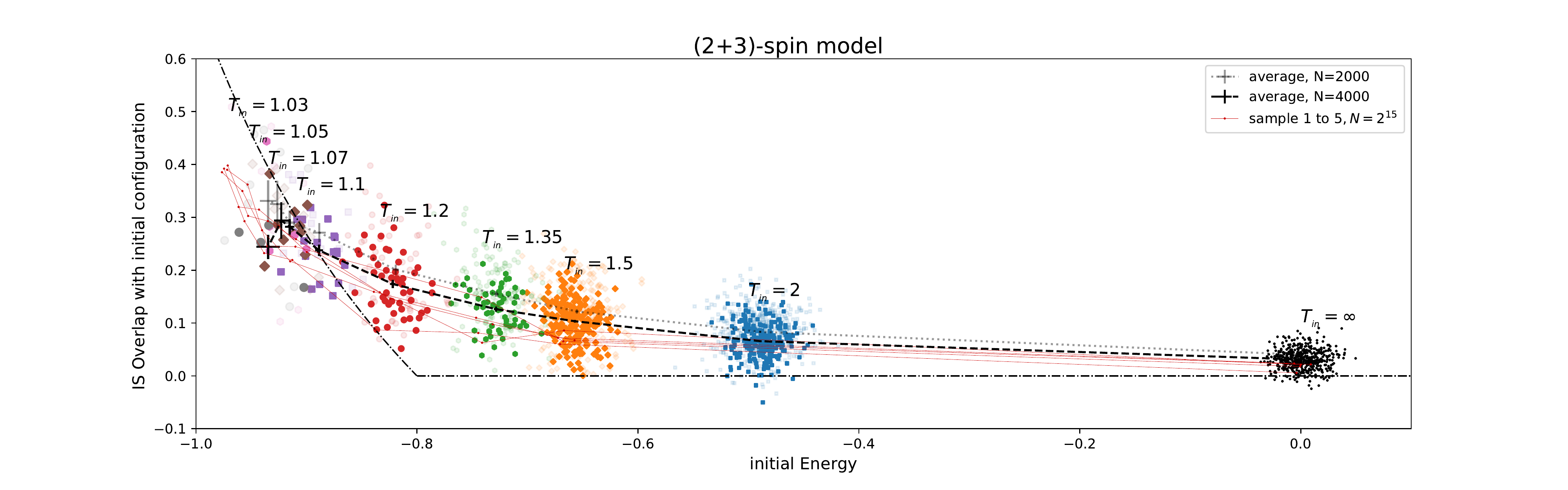}\\
	\vspace{-0.4cm}
	\includegraphics[width=0.9\columnwidth]{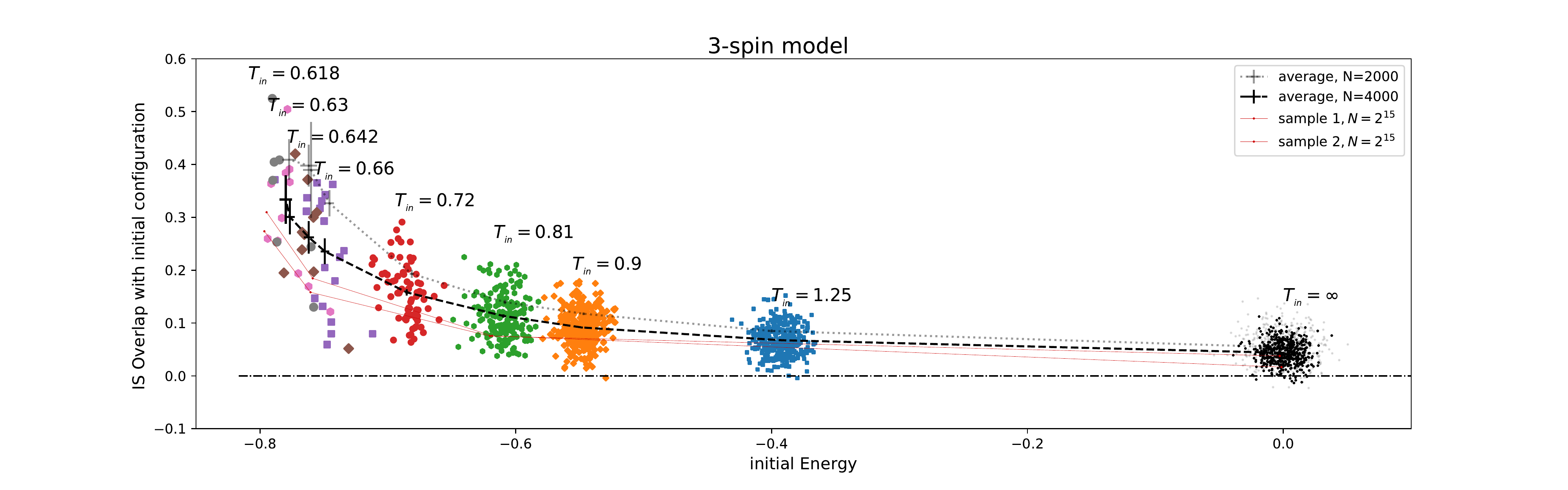}
	\caption{Overlap between the initial configuration and the IS final configuration for the same samples of size $N=4000$ presented in Figs.~\ref{fig:23spin} and \ref{fig:3spin}. The thin red lines report the overlaps measured in systems of size $N=2^{15}$ (each line is a different sample since we are forced to use planting for such a size). The dashed-dotted line is the $N\to\infty$ expectation that comes from the semi-empirical assumptions presented in \cite{folena_memories_2019}.}
	\label{fig:Corr}
\end{figure}
	
To conclude, we report in Fig.~\ref{fig:Corr} the overlap $q=\sum_{i}s_i(0)s^{\IS}_i/N$ between the inherent structure $s^{\IS}$ and the initial configuration $s(0)$ for both the ($2+3$)-spin and the $3$-spin models. The more the system is prepared near $\TMCT$, the more it keeps memory of the initial condition. However, while in pure models we expect such effect to vanish in the thermodynamic limit (if the WEB ansatz holds), in mixed models the situation may be more complicated according to the analysis in \cite{folena_memories_2019}: the system should keep memory of the initial configuration for every $T_{in}<\Tonset$ (SEB scenario).

For systems of sizes $N=2000,4000$ the difference between pure and mixed models is not evident. However, coming back to the simulation of planted systems of size $N=2^{15}$, and using the conjugated gradient dynamics to quickly find the ISs we get the results reported by thin red lines (one per sample, given we are using the planting trick). For the ($2+3$)-spin model the larger system presents almost the same behavior as the smaller ones at the lowest $T_{in}$, suggesting the data may be quite close to the thermodynamic limit. The dashed-dotted line is our best expectation in the thermodynamic limit obtained under some semi-empirical assumptions presented in \cite{folena_memories_2019}.

On the contrary, in the $3$-spin model the overlap in the large system gets rather suppressed with respect to the one in the smaller sizes, thus suggesting the overlap may vanish in the thermodynamic limit for all values of $T_{in}$ (thus recovering WEB).
The slow convergence toward zero of the overlap measured in the 3-spin model with $T_{in}$ close to $\TMCT$ does not come as a surprise. Already in the MFDE integration we observed such a very slow convergence when $T_{in}$ is close to $\TMCT$ \cite{folena_memories_2019}: in that case the slow convergence was observed as a function of time.

We conclude that the overlap is a poor observable ---in finite-size systems, as well as in the thermodynamic limit--- to spot the onset temperature (and the change from WEB to SEB), especially if compared to the energy which shows a sharper behavior around the onset.

\section{Conclusions}

We have simulated the gradient descent dynamics in the ($2+3$)-spin model which presents a RFOT, i.e.\ a mean-field model for structural glasses. Starting from configurations equilibrated near the mode coupling temperature $\TMCT$, we show that for large system sizes the dynamics follows closely the one predicted from the integration of the mean-field dynamical equations (MFDE) in the $N\to\infty$ limit. So we confirm the scenario proposed in Ref.~\cite{folena_memories_2019}: depending on the initial temperature $T_{in}$ the dynamics reaches different energies below the energy $E_{rc}$ reachable starting from a random condition. Moreover, we observe that for large enough systems the single gradient descent trajectory is self-averaging, i.e.\ for large $N$ it converges to the solution of the MFDE. 
	
We have analyzed the ISs reached by starting from an initial condition in equilibrium at temperature $T_{in}$, both in pure $3$-spin and mixed ($2+3$)-spin models. We have studied the correspondence between the initial and the final configurations of the gradient descent dynamics. We have defined the random initial energy $E_{rc}$ for finite size systems as the average IS energy reached from a random initial condition. Already for $N=2000$ and $N=4000$, in the ($2+3$)-spin, we observe an onset temperature $\Tonset(N)$, below which the average energy of ISs goes below $E_{rc}$. On the contrary, in the $3$-spin model, independently of the temperature of preparation, the system gets stuck (on average) on the threshold energy. 
	
From an algorithmic perspective, in mixed models ---contrary to pure ones--- different algorithms (e.g.\ cooling protocols) achieve different energies. Their lower-bound is the ‘algorithmic threshold' of the model.
It is not clear how to achieve such an algorithmic threshold, and not even how to compute it in general.
Recently it has been conjectured that in a particular class of mixed p-spin models\footnote{Models in this class present, at zero temperature, a spin glass phase with the replica symmetry broken infinitely many times (full-RSB) and the support of the overlap covering the whole range $q\in[0,1]$.} (to which none of the models studied in this work belong) an algorithm exists that achieves the optimal energy \cite{alaoui_algorithmic_2020,subag_free_2018}. However such an algorithm does not work on the space of configurations (i.e.\ on the $N$-dimensional sphere), but rather inside the sphere, building step-by-step the optimal configuration. So its comparison with the family of gradient descent algorithms that have access only to the information on the sphere of actual configuration is not totally fair. 
Given that algorithms working on the space of configurations (i.e.\ on the sphere) have a direct physical meaning, we believe it is an interesting open question to characterize their large time behavior both in the thermodynamic limit and for finite-size systems.
The results reported in Ref.~\cite{folena_memories_2019} and in the present work are the first steps in that direction.

Going back to the physical implications, the mixed p-spin gives us a new perspective on the organization of ISs in structural glasses. Sastry \textit{et al.} have proposed that, in glass-forming liquids, the onset in the slowing down of the dynamics is related to the structure of the energy landscape \cite{sastry_signatures_1998}. This correspondence was quantified by showing the ISs dependence on $T_{in}$ in the simulation of the glass-former Kob-Andersen model. So far there was no evidence that such a phenomenology could be observed in long-ranged (mean-field) models. The mixed $p$-spin model studied here is the first case. Recently an analogous ‘mean-field onset' has been observed in jamming of hard-spheres in the large dimensional limit \cite{charbonneau_memory_2020}. This support the idea that the onset follows from purely mean-field mechanisms.

In conclusion, mean-field models, despite their limitations, are capable to describe very complex mechanism in the physics of glass formers. The onset is one of them. So far we have demonstrated its existence, much more work is required to fully understand it.

\section{Acknowledgments}

We would like to thank Paolo Baldan and Stefano Sarao Mannelli for inspirational discussions. This research is supported by Simons Foundation Grants (No. 454941, S. Franz and No. 454949, G. Parisi); S. Franz is a member of the Institut Universitaire de France. 
	
\printbibliography
\newpage
	
\section*{Appendix A: Marginal Minima in Pure and Mixed Models}

While in \textit{pure} models $\TMCT$ has a role both in equilibrium and out-of-equilibrium dynamics, this is not the case in \textit{mixed} models where generally $\TSF<\TMCT$ and $\Tonset>\TMCT$. This simplification of the out-of-equilibrium dynamics in \textit{pure} models is due to the homogeneity of the Hamiltonian (\ref{homogeneous}), and can be understood by looking at the energy landscape of the model.
	
Let's start by the Hamiltonian $H[s]+\gamma/2 (\sum_{i}s_is_i-N)$, where $\gamma$ is the Lagrange multiplier to enforce the spherical constraint. We then define the extended gradient $\mathcal{G}$ and Hessian $\mathcal{H}$\footnote{note that both $\mathcal{G}$ and $\mathcal{H}$ are defined on the whole $\mathbb{R}^N$ into which the sphere is embedded}:
\begin{equation}\label{gradient_hessian}
	\mathcal{G}_{i}[s] \equiv H'_i[s] + \gamma s_i \qquad \mathcal{H}_{ij}[s]\equiv  H''_{ij}[s] + \gamma \delta_{ij}
\end{equation}
where $H'_i \equiv \nabla_{i} H[s]$ and $H''_{ij}[s] = \nabla_{i}\nabla_{j} H[s]$.  
Both $\mathcal{G}$ and $\mathcal{H}$ exhibit Gaussian fluctuations with means and variances\footnote{directly from the covariance defined in Eq. (\ref{fluctuations})}:
\begin{equation}
	\begin{aligned}
	\mathbb{E}[\mathcal{G}_{i}[s]] =  \gamma s_i &\qquad
	\mathbb{E}^{c}[\mathcal{G}_{i}[s] \mathcal{G}_{j}[\bar{s}]] = \delta_{ij}f'(\sum_{i}s_{i}\bar{s}_{i}/N)+O(1/N) \\ 
	\mathbb{E}[\mathcal{H}_{ij}[s]] =  \gamma \delta_{ij} &\qquad
	\mathbb{E}^{c}[\mathcal{H}_{ij}[s] \mathcal{H}_{kl}[\bar{s}]] = \frac{1}{N}(\delta_{ik}\delta_{jl}+\delta_{il}\delta_{jk})f''(\sum_{i}s_{i}\bar{s}_{i}/N)+O(1/N^2)  
	\end{aligned}
\end{equation}
At the leading order in $N$, fluctuations of the Hessian $\mathcal{H}$ are characteristic of matrices belonging to the Gaussian Orthogonal Ensemble and consequently present a Wigner semicircle-law spectrum with support:
\begin{equation}\label{support}
	[-2\sqrt{f''(1)}
	,2\sqrt{f''(1)}
	]
\end{equation}

The center of the spectrum is shifted by the Lagrange multiplier $\gamma$ that takes into account the local curvature of the sphere. This is the spectrum for each typical $\mathcal{H}$. 
	
A stationary point on the sphere is defined by the condition:
\begin{equation}
	P_i^k[s]\mathcal{G}_{k}[s] = 0 \qquad \Longrightarrow \qquad H'_{i}[s] - (\sum_{j}H'_{j}[s]s_j/N)s_i =0
\end{equation} 
where $P_i^k[s]=\delta_{ik}-s_is_k/N$ is the projector on the tangent plane to the sphere at $s$. Comparing with Eq.(\ref{gradient_hessian}), we see that stationary points have Lagrange multiplier equal to:
\begin{equation}
	\mu\equiv \gamma^* =  -\frac{\sum_{j}H'_{j}[s]s_j}{N}
\end{equation}
$\mu$ is called radial reaction and it gives the typical shift of the spectrum. From (\ref{support}) it follows that the condition to have a stable stationary point is:
\begin{equation}\label{marginality}
	\mu>\mu_{mg}\equiv 2\sqrt{f''(1)}
\end{equation}
Here, we have defined the marginal radial reaction $\mu_{mg}$.

In this analysis we have neglected rank-1 perturbation in the projected Hessian $\mathcal{H}^P$ that may give rise to isolated eigenvalues \cite{ros_complex_2019}. Calling $\Delta \mu = \mu-\mu_{mg}$, stable stationary points (minima) are the ones for which $\Delta \mu>0$.
In \textit{pure} models given the homogeneity (\ref{homogeneous}) we have that $\mu=- pE$, which implies that there exists only one value of energy for which the system is marginal $\Delta \mu=0$. In other words, the marginal manifold, i.e. the manifold which contains all the minima with $\Delta \mu=0$ lies entirely at the same energy:
\begin{equation}
	E_{th} \equiv -\mu_{mg}/p = -\sqrt{\frac{2(p-1)}{p}}
\end{equation}
In the thermodynamic limit ($N\to\infty$) this is the famous threshold energy, firstly introduced in \cite{cugliandolo_analytical_1993}. In \textit{pure} models, every dynamics starting from energies higher than $E_{th}$ will end up on this manifold. On the contrary in \textit{mixed} models, there is a whole range of energies for which there are marginal minima, and there is numerical evidence that taking different equilibrium temperatures $\TMCT<T <\Tonset$, the energy of the correspondent ISs are different. Consequently, it is not possible to define a threshold energy. Instead, we introduce the random initial condition energy $E_{rc}$ which can be defined in two different ways: \textit{statically} is the energy at which dominant minima become saddles (complexity calculations) and \textit{dynamically} is the energy reached in a quench from infinite temperature (asymptotic WEB ansatz). In the thermodynamic limit, these two definitions concord for any model chosen in the RFOT-class of $p$-spin models and give:
\begin{equation}\label{threshold}
	E_{rc} \equiv \frac{f(1) \left(f'(1)-f''(1)\right)-f'(1)^2}{f'(1) \sqrt{f''(1)}}
\end{equation}
whose derivation is reported in \cite{folena_memories_2019}. There, however, we used a different notation and $E_{th}$ (the threshold energy) refers ---both in pure and mixed models--- to the energy at which the dominant saddles become minima. Here we prefer to use the notation $E_{rc}$ (the random condition energy), whenever referring to general mixed models, in order to not mislead the intuition, since it does not correspond to an impassable threshold.  In finite size simulations we have considered the \textit{dynamical} definition of $E_{rc}$.
	
\section*{Appendix B: Finite Size Scaling of Inherent Structures}

Let's start by the simplest approximation. Let's admit that the gradient descent dynamics ends in the highest stable minima of the energy landscape. A minimum is stable as long as its lowest eigenvalue $\lambda_{min}$ is positive. In the thermodynamic limit ($N \to\infty$) this corresponds to the condition $\mu=\mu_{mg}$. However, in a system of finite size $N$, the lower edge presents fluctuations that scales as \cite{majumdar_top_2014}:
\begin{equation}
	\lambda_{min} \propto (\mu-\mu_{mg})N^{-2/3}
\end{equation}
The exponent $-2/3$ is directly connected to the shape of the Hessian spectrum, i.e. the Wigner semicircle $\rho(\lambda) \propto \sqrt{\mu_{mg}^2-(\lambda-\mu)^2}$. To understand this scaling we set the identity 
\begin{equation}
	\int_{\mu-\mu_{mg}}^{\mu-\mu_{mg}+\lambda_{min}} d\lambda \; \rho(\lambda) = 1/N
\end{equation}
which states that the probability of having at least one eigenvalue smaller of the smallest $\lambda_{min}$ should be proportional to $1/N$; and since the semicircle spectrum has a lower edge $\propto \lambda^{1/2}$, integrating we obtain the $-2/3$ scaling. Sampling a matrix from the GOE, the lowest eigenvalue fluctuates with $N$ according to:
\begin{equation}
	\lambda_{min} = \frac{\mu_{mg}}{2}\chi_{TW} N^{-2/3}
\end{equation}
where $\chi_{TW}$ is extracted from the Tracy-Widom distribution with $\beta=1$. This distribution has a non-zero average $\mathbb{E}[\chi_{TW}]\approx-1.2$, so we expect that in average, a minimum has a marginal radial reaction smaller than the one in the thermodynamic limit:
\begin{equation}
	\mu_{mg}(N) \equiv \mu_{mg}(1+\frac{1}{2}\mathbb{E}[\chi_{TW}] N^{-2/3}) < \mu_{mg}
\end{equation}

\begin{figure}[ht]
	\centering
	\includegraphics[width=0.45\columnwidth]{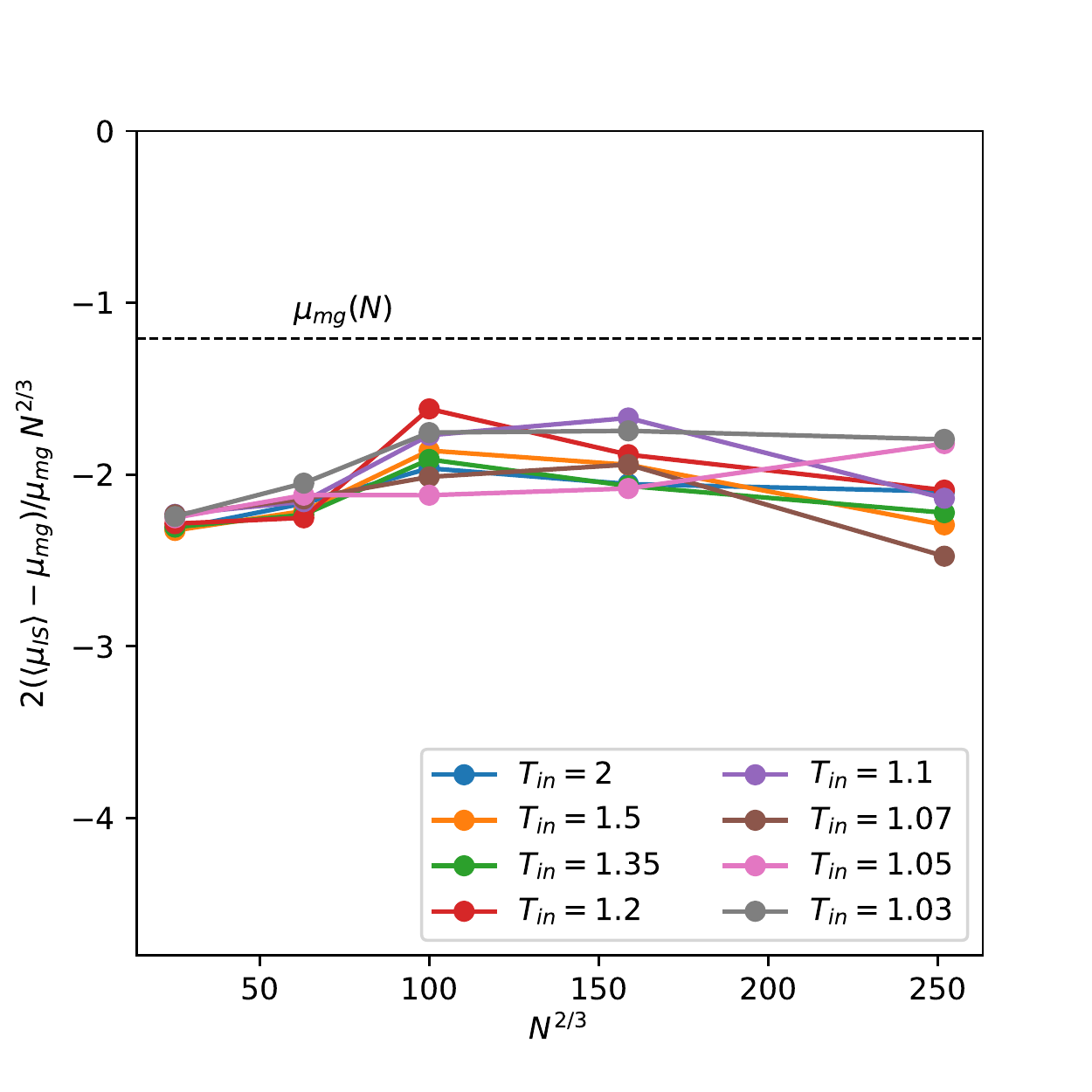}
	\hfill
	\includegraphics[width=0.45\columnwidth]{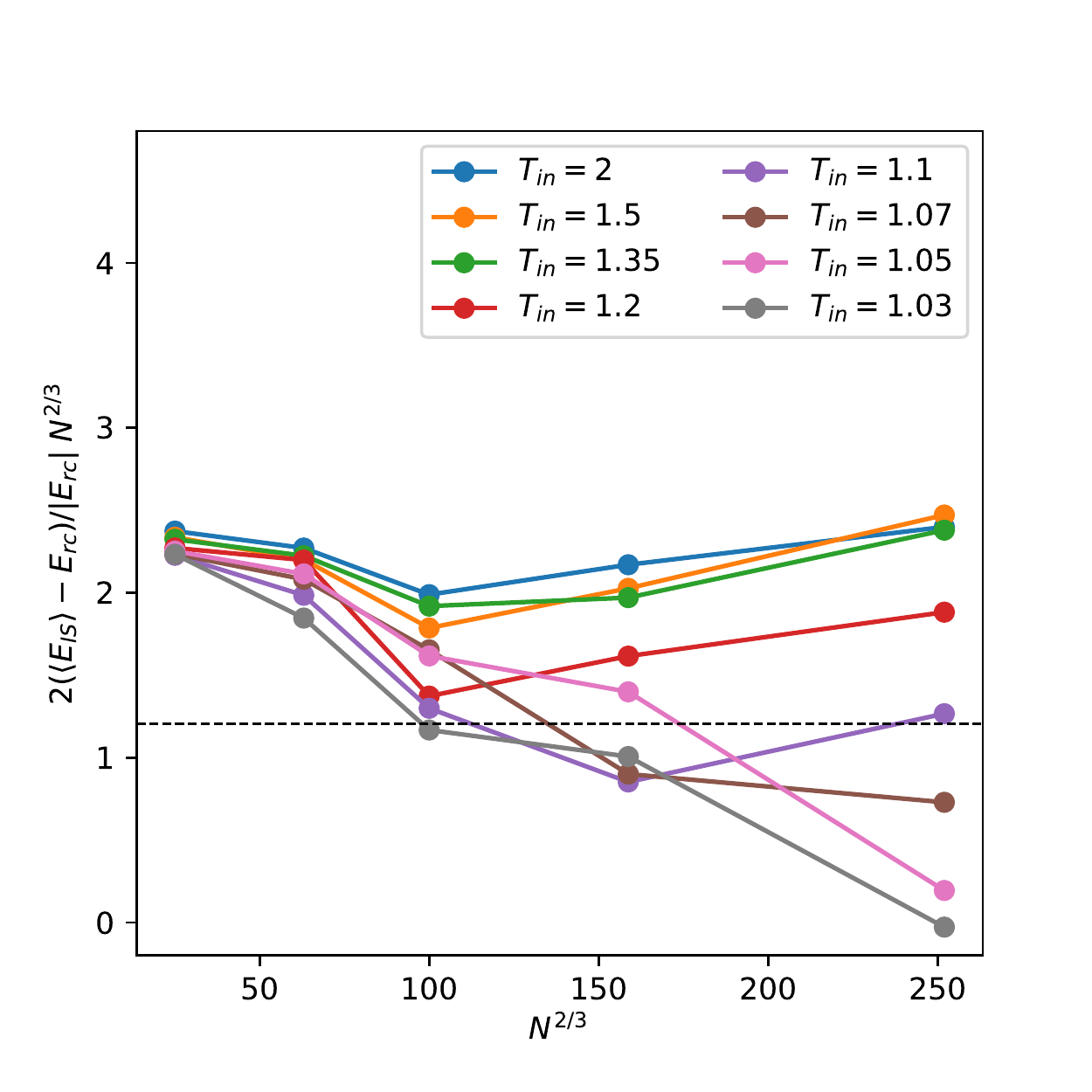}\\
	\caption{\footnotesize Finite size scaling of ISs in the $(2+3)$-spin model. The ISs are obtained through conjugated gradient descent, starting from equilibrium configurations sampled at different temperatures $T_{in} = 2,1.5,1.35,1.2,1.1,1.07,1.05,1.03$, greater than $\TMCT$ (as described in \ref{onset}). For each size $N=125,250,500,1000,2000,4000$ a unique system is considered. \textbf{Left:} the average of the radial reaction $\langle\mu_{IS}\rangle$ over different ISs rescaled around its thermodynamic limit $\mu_{mg}$. We notice that it is almost independent from the temperature $T_{in}$.  \textbf{Right:} the same rescaling for the average IS energy $\langle E_{IS}\rangle$ around the random condition $E_{rc}$ has a strong dependence on the temperature for $T_{in}<1.35$.}
	\label{E_mu_mg}
\end{figure}

This same behavior is observed when looking at inherent structure of the $(2+3)$-spin model reached from different temperatures. In Fig.\ref{E_mu_mg}:left is shown the average radial reaction of the inherent structures $\langle\mu_{\IS}\rangle$ reached from different temperatures; and as expected from the above analysis, it is lower than that of the thermodynamic limit $\mu_{mg}$. 

\begin{figure}[ht]
	\hspace*{-2cm}\includegraphics[width=1.2\columnwidth]{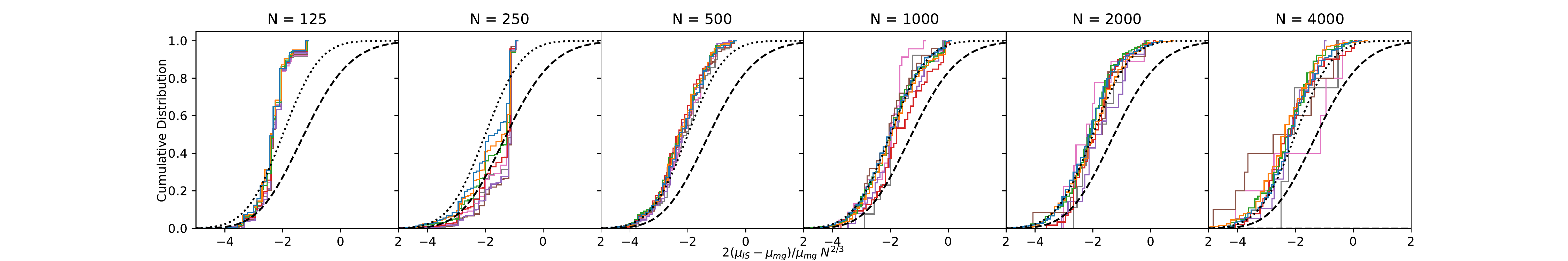}
	\hspace*{-2cm}\includegraphics[width=1.2\columnwidth]{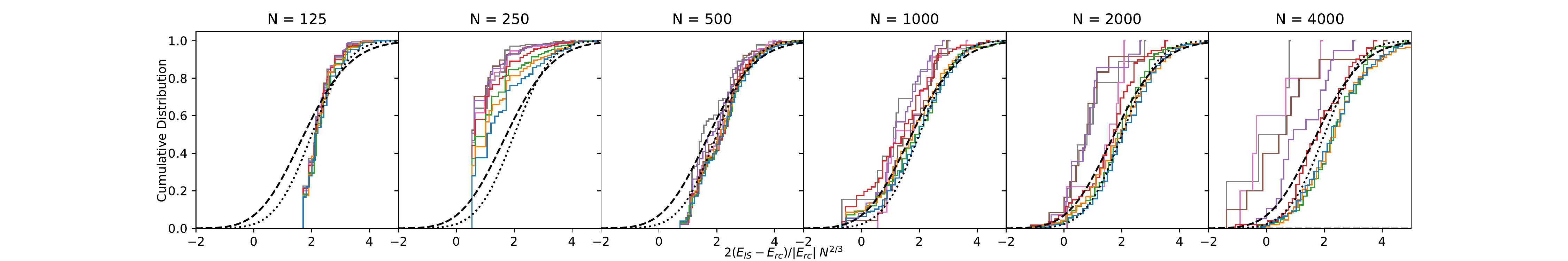}
	\caption{\footnotesize Rescaled cumulative distribution of IS radial reaction and energy for the same data presented in Fig. \ref{E_mu_mg}. \textbf{Top:} cumulative distribution of the radial reaction $\mu_{\IS}$. The Tracy-Widom cumulative expected from a “naive" GOE analysis, not considering the dynamics, is plotted with a dashed black line. The Gaussian cumulative distribution with variance 1 and mean $\pm 2$ is shown with a dotted black lines. A similar shape is observed in all systems greater than $N=500$. \textbf{Bottom:} cumulative distribution of the energy $E_{\IS}$. Here, the rescaling is always centered around $E_{rc}$ so for smaller temperatures the distribution is shifted to the left.  }
	\label{E_mu_TW}
\end{figure}

However, the IS minima are not random but reached through dynamics, and therefore, the resulting rescaled distribution is not Tracy-Widom \cite{boltz_fluctuation_2019}. This can be observed in the top plot of Fig. \ref{E_mu_TW}, where the rescaled cumulative distribution of radial reaction $\mu_{\IS}$ for different sizes and temperatures is compared with the Tracy-Widom cumulative distribution (dashed black lines).

The same scaling $N^{-2/3}$ can be seen in the energy of the inherent structures $E_{\IS}$. However, in this case the thermodynamic value depends on the temperature. So doing the rescaling around $E_{rc}$ we observe (Fig. \ref{E_mu_mg}:right) a spreading of the different curves increasing the size of the system. Just for comparison, we also show the rescaled cumulative distributions for $E_{\IS}$ in Fig.\ref{E_mu_TW}:bottom.

It is important to note that contrary to the energies $E_{\IS}$, the radial reactions $\mu_{\IS}$ seem to have the same average value, independently of the temperature $T_{in}$, and the same rescaled distribution independently of the size of the system $N$  (for $N>500$). This is directly connected to the main role of this parameter, which sets the stability of the system, defining the location of the lower edge of Hessian spectrum. 
	
\begin{figure}[ht]
	\hspace*{-2cm}     
	\includegraphics[width=1.2\columnwidth]{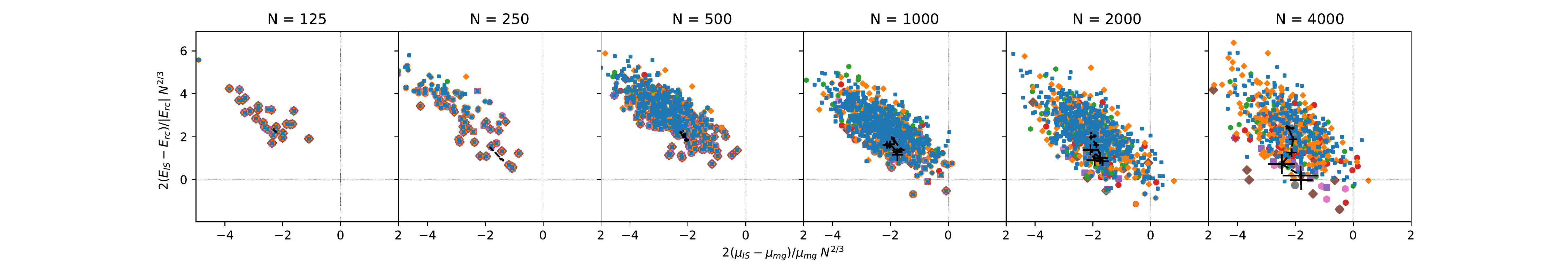}
	\caption{\footnotesize Scatter plot of radial reaction $\mu_{\IS}$ and energy $E_{\IS}$ of the ISs reached from different temperatures and for different system sizes. Again, the same data presented in Fig. \ref{E_mu_TW} and Fig. \ref{E_mu_mg}. The scatter plots are rescaled with the expected $N^{-2/3}$ power. The black lines are the averages at every temperature. Different colors represent ISs reached from different temperatures. The systems of sizes $N=125$ and $N=250$ seem too small for studying the large finite size scaling of the model. }
	\label{scatter}
\end{figure}

Finally in Fig. \ref{scatter} is reported the rescaled scatter plot of both energy and radial reaction. We see that both system $N=125$ and $N=250$ present a small number of IS minima which can be reached from any temperature. These sizes are too small for studying the $(2+3)$-spin model. 

\end{document}